\documentclass[a4paper,fleqn,usenatbib]{mnras}

\usepackage{newtxtext,newtxmath}

\usepackage[T1]{fontenc}
\usepackage{ae,aecompl}

\usepackage{graphicx}	
\usepackage{amsmath}	
\usepackage{amssymb}	

\title[Comet 67P formed through SI/GI]{Evidence for the formation of comet 67P/Churyumov-Gerasimenko through gravitational collapse of a bound clump of pebbles}

\author[J. Blum et al.]{J\"urgen Blum,$^{1}$\thanks{E-mail: j.blum@tu-bs.de}
Bastian Gundlach,$^{1}$
Maya Krause,$^{1}$
Marco Fulle,$^{2}$
\newauthor
Anders Johansen,$^{3}$
Jessica Agarwal,$^{4}$
Ingo von Borstel,$^{1}$
Xian Shi,$^{4}$
Xuanyu Hu,$^{4,1}$
\newauthor
Mark S. Bentley,$^{5}$
Fabrizio Capaccioni,$^{6}$
Luigi Colangeli,$^{7}$
Vincenzo Della Corte,$^{6}$
\newauthor
Nicolas Fougere,$^{8}$
Simon F. Green,$^{9}$
Stavro Ivanovski,$^{6}$
Thurid Mannel,$^{5,10}$
\newauthor
Sihane Merouane,$^{4}$
Alessandra Migliorini,$^{6}$
Alessandra Rotundi,$^{11,6}$
\newauthor
Roland Schmied,$^{5}$ and
Colin Snodgrass,${}^{9}$
\\
$^{1}$Institut f\"{u}r Geophysik und extraterrestrische Physik, Technische Universit\"{a}t Braunschweig, Mendelssohnstr. 3, 38106 Braunschweig, Germany\\
$^{2}$INAF -- Osservatorio Astronomico, Via Tiepolo 11, 34143 Trieste, Italy\\
$^{3}$Lund Observatory, Lund University, S\"{o}lvegatan 27, 223 62 Lund, Sweden\\
$^{4}$Max-Planck-Institut f\"{u}r Sonnensystemforschung, Justus-von-Liebig-Weg 3, 37077 G\"{o}ttingen, Germany\\
$^{5}$Space Research Institute, Austrian Academy of Sciences, Schmiedlstrasse 6, 8042 Graz, Austria\\
$^{6}$INAF -- Istituto di Astrofisica e Planetologia Spaziali, Via Fosso del Cavaliere 100, 00133 Rome, Italy\\
$^{7}$ESA - ESTEC, European Space Agency, Keplerlaan 1, 2201 AZ Noordwijk, The Netherlands\\
$^{8}$Department of Climate and Space Sciences and Engineering, University of Michigan, Ann Arbor, MI 48109, USA\\
$^{9}$Planetary and Space Sciences, School of Physical Sciences, The Open University, Milton Keynes MK7 6AA, UK\\
$^{10}$University of Graz, Universit\"{a}tsplatz 3, 8010 Graz, Austria\\
$^{11}$Universit\'{a} degli Studi di Napoli Parthenope, Dip. di Scienze e Tecnologie, CDN IC4, 80143 Naples, Italy
}

\date{Accepted 2017 October 17. Received 2017 October 17; in original form 2017 Feburary 3
}

\pubyear{2015}

\begin{document}
\label{firstpage}
\pagerange{\pageref{firstpage}--\pageref{lastpage}}
\maketitle

\begin{abstract}
The processes that led to the formation of the planetary bodies in the Solar System are still not fully understood. Using the results obtained with the comprehensive suite of instruments on-board ESA's Rosetta mission, we present evidence that comet 67P/Churyumov-Gerasimenko likely formed through the gentle gravitational collapse of a bound clump of mm-sized dust aggregates (``pebbles''), intermixed with microscopic ice particles. This formation scenario leads to a cometary make-up that is simultaneously compatible with the global porosity, homogeneity, tensile strength, thermal inertia, vertical temperature profiles, sizes and porosities of emitted dust, and the steep increase in water-vapour production rate with decreasing heliocentric distance, measured by the instruments on-board the Rosetta spacecraft and the Philae lander. Our findings suggest that the pebbles observed to be abundant in protoplanetary discs around young stars provide the building material for comets and other minor bodies.
\end{abstract}




\section{Introduction}

Comets are thought to be primitive, because, owing to their small size, the amount of lithostatic compression and thermal metamorphism they experienced before they entered the inner Solar System is smaller than for larger bodies. For most of their life since formation they orbited the Sun at such large distances that they have remained almost unaffected by solar irradiation. Thus, comets are the perfect witnesses to the processes that led to the formation of the planetesimals, the building blocks of the planets.

Due to the extended period of investigation of comet 67P/Churyumov-Gerasimenko (hereafter 67P) by the Rosetta spacecraft, which encompassed heliocentric distances between 3.6 au and 1.2 au pre- and post-perihelion, and its comprehensive suite of experiments, our knowledge of 67P, and generally about comets, has increased enormously. For brevity, we refer to the Rosetta orbiter and Philae lander simply as Rosetta. The set of data delivered by Rosetta allows us to decipher the formation processes from $\sim \mu$m-sized dust and ice particles to bodies with sizes of a few kilometres across. Simultaneously, the formation model will be used to derive measurable properties of active comets. These observables encompass structural parameters, such as size, porosity and homogeneity of the comet nucleus, as well as properties related to the activity of 67P, like outgassing rate and dust activity. \citet{davidsson2016} have recently derived a comprehensive model for the formation of the bodies in the Kuiper Belt, which is based on the calculations of \citet{weidenschilling1997} who predicted that cometesimals grow smoothly from microscopic particles all the way to kilometre sizes through accretion of bodies that are typically a factor of a few smaller in size than themselves. Thus, in this model there is no single characteristic size scale between the monomer grains and the cometesimal itself. In the following, we will show that the specific properties of 67P imply that it formed by the gentle gravitational collapse of a bound clump of mm-sized dust aggregates (``pebbles''), which survived to the present day, in contrast to the growth model by \citet{weidenschilling1997}.

\section{Planetesimal formation model}
The current understanding of how planetesimals form in a protoplanetary disc (PPD) is that initially all collisions among the solid dust or ice particles lead to sticking. This is because the van der Waals force \citep{heim1999} is strong enough to bind the grains together when they initially collide at very low speeds ($\ll 1\ \mathrm{m\ }{\mathrm{s}}^{-1}$, \citet{weidenschilling1977a}). This hit-and-stick process leads to the formation of fractal aggregates \citep{blum2000a,krause2004,fulle2017b}. If the solid particles are either non-icy or larger than 0.1 $\rm \mu$m in size \citep{kataoka2013,lorek2017}, fractal growth comes to a halt, when the impact energy exceeds the restructuring limit \citep{dominik1997,blum2000b}. However, mutual collisions still lead to the growth of the aggregates until the bouncing barrier is reached for aggregate sizes of $\sim \mathrm{1\ cm}$ \citep{zsom2010} in an assumed minimum-mass solar nebula  (MMSN) model \citep{weidenschilling1977b} at 1 au, and $\sim \mathrm{1\ mm}$ in the comet-formation regions, respectively \citep{lorek2017}, and perhaps larger for somewhat ``stickier'' organic materials if the temperatures are $\sim 250\ \mathrm{K}$ \citep{kudo2002}. The growth times to the bouncing barrier are consistently $\sim 1000$ orbital time-scales \citep{lorek2017}. This is within the lifetime of the solar nebula of a few million years \citep{ribas2015}. Continued bouncing then leads to the rounding and compaction of the aggregates whose filling factor increases from $\phi_{\mathrm{pebble}}\approx 0.05$ to $\phi_{\mathrm{pebble}}\approx 0.40$ \citep{weidling2009}. The filling factor of the aggregates describes the fraction of the pebble volume actually occupied by matter. The collision processes described here have been extensively investigated in the laboratory and can be considered robust \citep{blum2008,guettler2010}. The behaviour described above is valid for refractory materials but may change significantly for water ice with its tenfold increased sticking threshold \citep{kataoka2013,gundlach2015}. However, observations imply that the composition of 67P is dominated by non-volatile materials \citep{fulle2016b,capaccioni2015,quirico2016}.

\begin{figure}
	\includegraphics[width=\columnwidth]{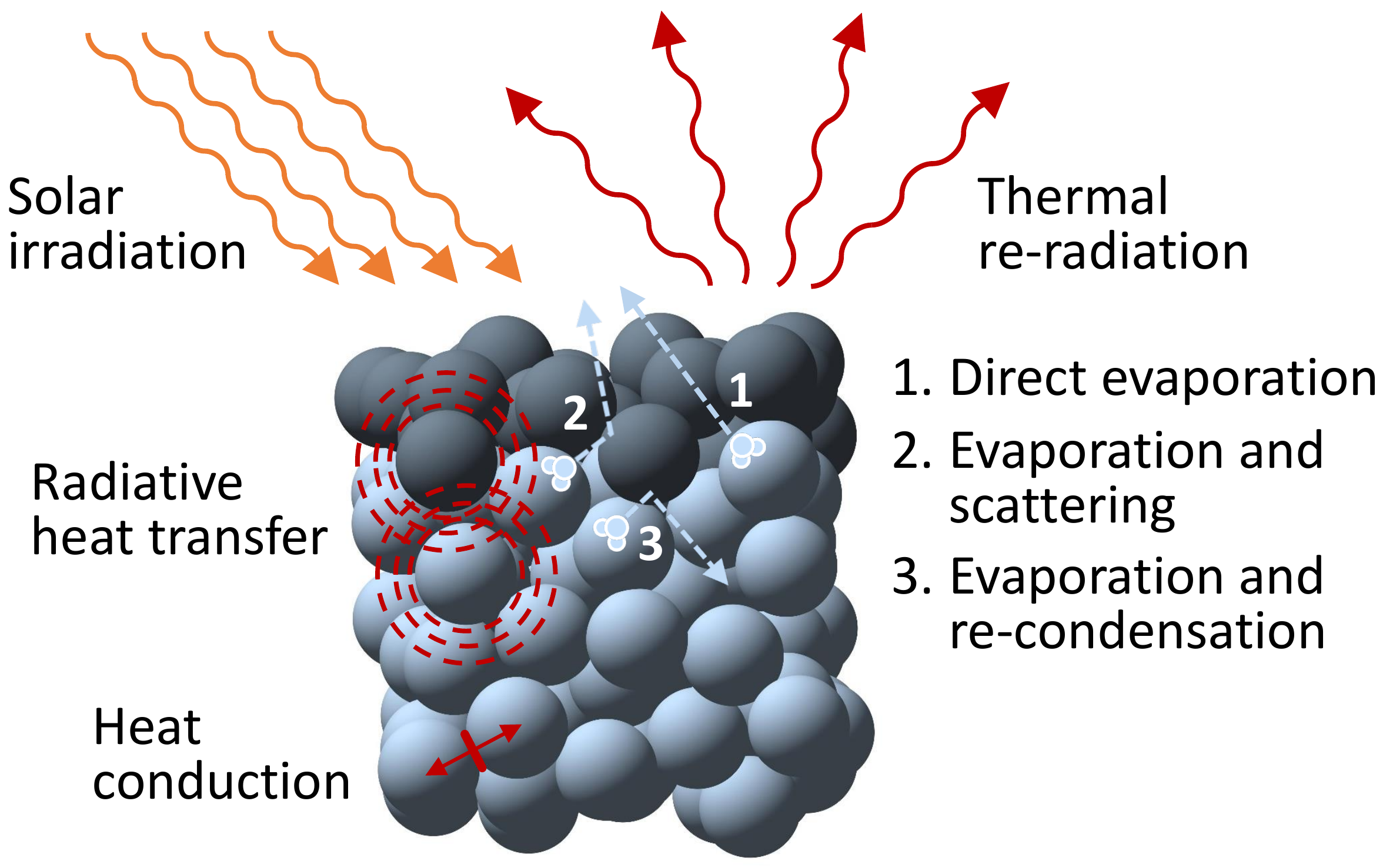}
    \caption{Schematic of the structure of comet nuclei. If comet nuclei formed by gentle gravitational collapse of a bound clump of dust aggregates, their packing morphology should be close to what is depicted in the figure. The dust pebbles are here shown as spheres and have a packing fraction of $\phi \approx 0.6$. At formation, the pebbles contain refractory materials and ices (light grey spheres). When approaching the Sun, pebbles close to the surface of the comet nucleus lose their icy constituents (dark grey spheres). Here, we assume that all pebbles whose centres are within two pebble radii above the ice line become desiccated. The evaporation of water molecules can either happen directly into vacuum (dashed arrow indexed 1), through a scattering process (2), or into the interior, where re-condensation occurs (3). The major energy flows are also indicated: absorption of solar radiation at the surface, re-radiation of heat from the surface, energy transport through the solid contacts between pebbles, and thermal radiation in the interior of the comet nucleus, respectively.}
    \label{fig:01}
\end{figure}

The next step in planetesimal formation is more controversial. We argue that planetesimal formation proceeded through spatial concentration of the mm-sized dust pebbles by the streaming instability \citep{youdin2005} that ultimately led to the formation of a gravitationally bound cloud of pebbles, which gently ($\lesssim 1\ \mathrm{m\ }{\mathrm{s}}^{-1}$) collapsed to form a planetesimal (\citet{johansen2007}; see Figure \ref{fig:01}). The streaming instability is a collective effect of dust particles. An individual dust particle with Stokes number $St \ll 1$ is suspended in the gas, which forces it to travel with sub-keplerian velocity. This, in turn, leads to a rapid radial drift inward \citep{weidenschilling1977a}. In contrast, a group of such grains is less affected and, thus, moves with a higher orbital velocity, which suppresses the radial drift. Here, the Stokes number is defined by $St=t_{\mathrm{f}}\Omega_k$, with $t_{\mathrm{f}}$ and $\Omega_k$ being the particle stopping time in the gas and the Kepler frequency of the particle's orbit, respectively. If the concentration of the particles is sufficiently large, their feedback to the gas even strengthens the effect. Thus, such an instability region attracts all particles that cross its orbit, which leads to a rapid growth in mass concentration. If this concentration is high enough, the bound pebble cloud can gravitationally collapse to form a granular body \citep{johansen2007}. It was shown that the dust-to-gas ratio\footnote{If we assume that in the comet-forming regions the temperatures were so low that basically all elements heavier than Helium were condensed, this value equals the ``metallicity''. In this context, ``dust'' also encompasses the condensed volatiles, i.e., the ices, while ``gas'' consists of $\mathrm{H}_2$ and He, with only minor contributions of other species.} of the PPD plays an important role for the onset of the streaming instability. A slightly increased dust-to-gas ratio is required, while the streaming instability does not occur for solar metallicity \citep{carrera2015,yang2016}. This means that planetesimal formation through the streaming instability must be delayed until part of the gaseous PPD has been dissipated. The disc lifetime is a few million years \citep{ribas2015} so that such a delayed planetesimal formation would also help to prevent the planetesimals from melting by short-lifetime radioactive decay \citep{davidsson2016}.

The detection of extremely fluffy dust particles with filling factors $\lesssim 0.001$ (comprising $\sim$1\% of the total mass, but consequently a much larger share of the volume of the comet nucleus) by the Rosetta/GIADA instrument \citep{fulle2015} and the discovery of a fractal particle by the Rosetta/MIDAS instrument \citep{mannel2016} unambiguously show that the comet never experienced any global compression, because otherwise, the maximum yield strength of the fluffy aggregates would have been overcome and the aggregates compacted to filling factors $\phi_{\mathrm{pebble}} \approx 0.4$. The maximum dynamic yield strength can be estimated following the numerical model by \citet{dominik1997} and the experimental results by \citet{blum2000b} and is $Y=\rho v_{\mathrm{mc}}^2 /2 \approx 1$ Pa, with $v_{\mathrm{mc}} \approx 1 \mathrm{\ m \ s^{-1}}$ and
$\rho \approx 1 \mathrm{\ kg \ m^{-3}}$ being the impact velocity leading to maximum compaction \citep{dominik1997,blum2000b} and the mass density of the fractal aggregates, respectively.

A statistical analysis of the occurrence of showers of fluffy particles in the GIADA instrument shows that the fluffy parent aggregates must have had sizes on the order of a few mm \citep{fulle2017b}. To fit these aggregates into the void spaces between the pebbles, the latter must have (average) radii of $a=8.5\ \mathrm{mm}$ \citep{fulle2017b}. It should be mentioned that the porous dust aggregates detected by MIDAS and GIADA possess a fractal structure with consistent fractal dimensions, which establishes the co-existence of a fluffy/fractal and a pebble population \citep{fulle2017b}. These fluffy aggregates must be intact remnants of the initial fractal dust agglomeration before impact compaction \citep{fulle2017b}, as suggested by \citet{mannel2016}. Due to their slow relative velocities in the solar nebula, on the order of a few cm s${}^{-1}$ \citep{weidenschilling1977a}, collisions between the low-mass fluffy aggregates and the cm-sized pebbles result in sticking, without any significant compaction of the fluffy particles \citep{guettler2010,blum2008,weidling2009}. The survival of fluffy dust aggregates until today is only possible if they were encased in the void space between the pebbles and if the pebbles survived the formation process of the nucleus of 67P intact. The presence of fractal aggregates rules out all formation processes other than a gentle gravitational collapse of a bound clump of pebbles \citep{fulle2017b}. In the alternative option, collisional coagulation, favoured by \citet{davidsson2016}, high-porosity or fractal dust aggregates would have experienced destructive high-speed (up to $\sim 50\ \mathrm{m\ }{\mathrm{s}}^{-1}$) collisions \citep{weidenschilling1977a}. In the following, we will show that the pebbles forming comet 67P have radii between $\sim 1$ mm and $\sim 6$ mm, thus confirming that the presence of fractal dust in 67P requires a real change of paradigm regarding the collisional history of the outer Solar System \citep{farinella1996,morbidelli2015,fulle2017b}.

The question remains whether 67P was formed as we observe it today or through re-accretion of fragments from a catastrophic collision of an initially larger body \citep{morbidelli2015,rickma2015}. The fact that the pebbles stayed intact shows that the planetesimal that later became comet 67P had a radius of $<50\ \mathrm{km}$, because otherwise the collapse would have destroyed the pebbles \citep{bukhari2017,wahlberg2017}. Recent numerical simulations indicate that a catastrophic collision between two pebble-pile objects can lead to the gravitational re-accretion of a km-sized body, preferentially from uncompressed and unheated material from the target body \citep{michel2016}. Once the spatial resolution in such simulations is fine enough to follow the fate of individual pebbles, it may be decided whether or not such a secondary formation scenario of comet nuclei is in agreement with the presence of fractal dust aggregates.

\section{The size of the pebbles forming comet 67P}
\begin{figure}
	\includegraphics[width=\columnwidth,angle=90]{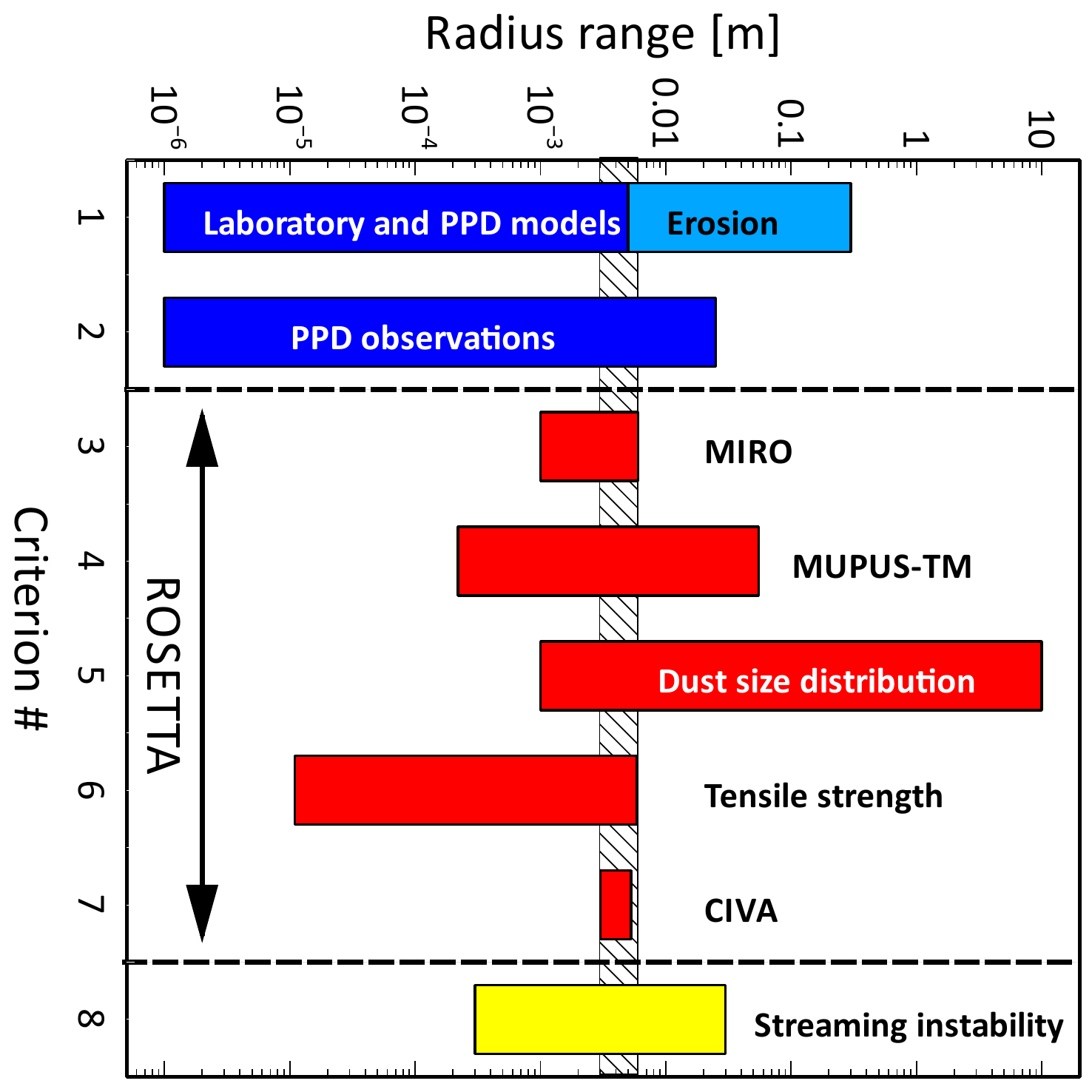}
    \caption{Size ranges of dust aggregates. From left to right: protoplanetary-disc models and observations (blue, criteria 1-2), Rosetta observations (red, criteria 3-7), and streaming instability criterion (yellow, criterion 8). The hatched region is the minimum width for pebble radii consistent with all Rosetta observations.}
    \label{fig:02}
\end{figure}

Below, we list arguments consistent with the pebble hypothesis of planetesimal formation. We will start with empirical, astronomical and theoretical aspects (criteria 1 and 2 below), followed by Rosetta observations of comet 67P (criteria 3-7), and finally will come back to the streaming instability criterion (criterion 8). In Figure \ref{fig:02}, we show ranges for the radii of the dust pebbles, either observed or theoretically predicted, for each of the eight items.

\subsection{Protoplanetary-disc models and observations}

\textit{1. Pebble-size constraint from laboratory experiments and protoplanetary-disc models.} From the standpoint of numerical disc simulations and laboratory data for the collisional behaviour of dust and ice aggregates, there are two obstacles that cannot be overcome. The first is the fragmentation barrier (dark blue in Figure \ref{fig:02}). When two dust aggregates collide above a certain impact velocity, disintegration of the dusty bodies is so strong that the largest surviving fragment has less mass than the original target \citep{bukhari2017}. In typical PPDs and outside dust-dominated sub-discs (where the velocities are not caused by dust-gas interactions and are, thus, very different), this means that collisions among bodies with size ratios $\le 5$ lead to the fragmentation of both bodies \citep{bukhari2017} if they exceed $\sim 1$ cm in diameter for dust- and $\sim 10$ cm for ice-dominated materials, based upon velocities given by \citet{weidenschilling1997}. The second obstacle is the erosion barrier (light blue in Figure \ref{fig:02}). Even if some ``lucky survivors'' could escape catastrophic fragmentation, they cannot avoid being hit by small particles and agglomerates at relatively high velocities (exceptions are, again, dust-dominated sub-discs in turbulence-free dead zones). Experiments show that impacts by individual dust grains or small aggregates lead to mass loss of the target aggregate \citep{schraepler2011,krijt2015,schraepler2017}, which limits the radii to $\lesssim 30\ \mathrm{cm}$ for dust agglomerates in the region of 10-100 au. As the impact process liberates more fragments in the relevant size range than it consumes, erosion is inevitable.

Monte Carlo simulations using realistic collision physics show that the maximum aggregate size that can be achieved by direct coagulation depends on the heliocentric distance and monomer-grain size. \citet{lorek2017} derived maximum aggregates masses between $\sim 0.1$ g at 5 au (only slightly dependent on monomer size) and $\sim 10^{-5}-10^{-3}$ g at 50 au for 1-0.1 $\rm \mu m$ monomer size. With the fragmentation and erosion barriers (see above) in mind, we conclude that the maximum dust-aggregate diameter cannot exceed $\sim 1$ cm. \citet{schraepler2017} showed that the interplay between coagulation and erosion leads to a non-negligible mass fraction in very small grains and agglomerates so that we set the minimum dust size in Figure \ref{fig:02} to 1 $\rm \mu m$.

\textit{2. Pebble-size constraint from observations of protoplanetary discs.}

Over the past decades, observations of PPDs have been performed over a wide range of wavelengths \citep[see, e.g., the reviews by][]{williams2011,testi2014}. Visible and near-infrared observations of mostly scattered stellar light are sensitive to $\rm \mu m$-sized dust and show the presence of these particles in PPDs of all ages \citep[see, e.g., the discussion in][]{dullemeond2005}. At longer wavelengths, thermal emission of large dust aggregates from close to the midplane dominates over scattering in the disc's atmosphere. Very-long-wavelength observations show the presence of mm-sized and larger grains in PPDs \citep[see, e.g., ][]{vanboekel2004,dalessio2006,natta2007,birnstiel2010,ricci2010,perez2012}.

At the current stage, it is unclear which aggregate sizes dominate the dust-mass spectrum of PPDS. However, there seems to be a size cut-off above which the abundance of larger grains considerably drops. Recently, spatially resolved observations of PPDs could determine this cut-off size for a range of distances to the central star. \citet{testi2014} report that typically the size of the dust-emission region becomes smaller with increasing observation wavelength, which indicates that grain growth in the inner disc leads to larger grains than further out, in qualitative agreement with the model by \citet{lorek2017}. \citet{testi2014} show two examples where mm-sized aggregates prevail at distances of 50-100 au, whereas around 20 au centimetre particles can be found. Recently, \citet{tazzari2016} performed interferometric observations of three protoplanetary discs \citep[including one source from the][study]{testi2014} at wavelengths of 0.88 to 9.83 mm and derived the dust size cut-off of otherwise continuous size distributions at high spatial resolution. Similar to the two PPDs reported by \citet{testi2014}, \citet{tazzari2016} found that the maximum dust-aggregate radii decreased with increasing distance to the central star: at 10 au, the pebble sizes are around 25 mm, whereas at 50 au, the maximum dust radii reach only 1-5 mm. \citet{tazzari2016} analysed that the maximum size falls off with distance to the central star following a power law with exponent $\sim$ -1.5, whereas the Monte-Carlo simulations by \citet{lorek2017} find exponents $\gtrsim -1$. Dust properties at various distances to the central star have also been determined by \citet{trotta2013,perez2015,liu2017}, which confirm grain growth with a maximum grain size decreasing with increasing distance.

These recent findings clearly prove the existence of pebble-sized dust aggregates at comet-formation distances. The maximum pebble size found by \citet{tazzari2016} is shown in Figure \ref{fig:02} as the maximum of the dust-size distribution and as a proxy for the many other measurements. Here, the full range of observed dust sizes in PPDs from micrometres to centimetres is displayed.

\subsection{\label{Sect:RosettaData} Observations of comet 67P by Rosetta/Philae instruments}

\textit{3. Pebble-size constraint from measurements of the sub-surface temperatures by MIRO.} The MIRO instrument was the only sensor on-board the Rosetta orbiter to measure the subsurface temperature of the cometary nucleus. MIRO is a microwave radiometer in two narrow sub-millimetre and millimetre wavelength bands at ${\lambda}_{\mathrm{sub-mm}}=\ 0.533\ \mathrm{mm}$ and ${\lambda}_{\mathrm{mm}}=\ 1.594\ \mathrm{mm}$, respectively, with a relative bandwidth of $\Delta \lambda / \lambda = 2 \times 10^{-3}$  \citep{schloerb2015,gulkis2015}. Thus, MIRO measures the thermal fluxes emitted by the cometary material in the Rayleigh-Jeans wavelength regime. Typical penetration depths are up to a few centimetres, depending on the absorption efficiency of the cometary material at the two MIRO wavelengths. We applied a thermophysical model (see Appendix A) to a synthetic comet consisting of pebbles and determined the diurnal temperature evolution at depths of up to 50 cm for a wide range of pebble radii and absorption coefficients for a position of 25$^\circ$ northern latitude and a heliocentric distance of 3.27 au.

\begin{figure*}
   	\includegraphics[width=8cm]{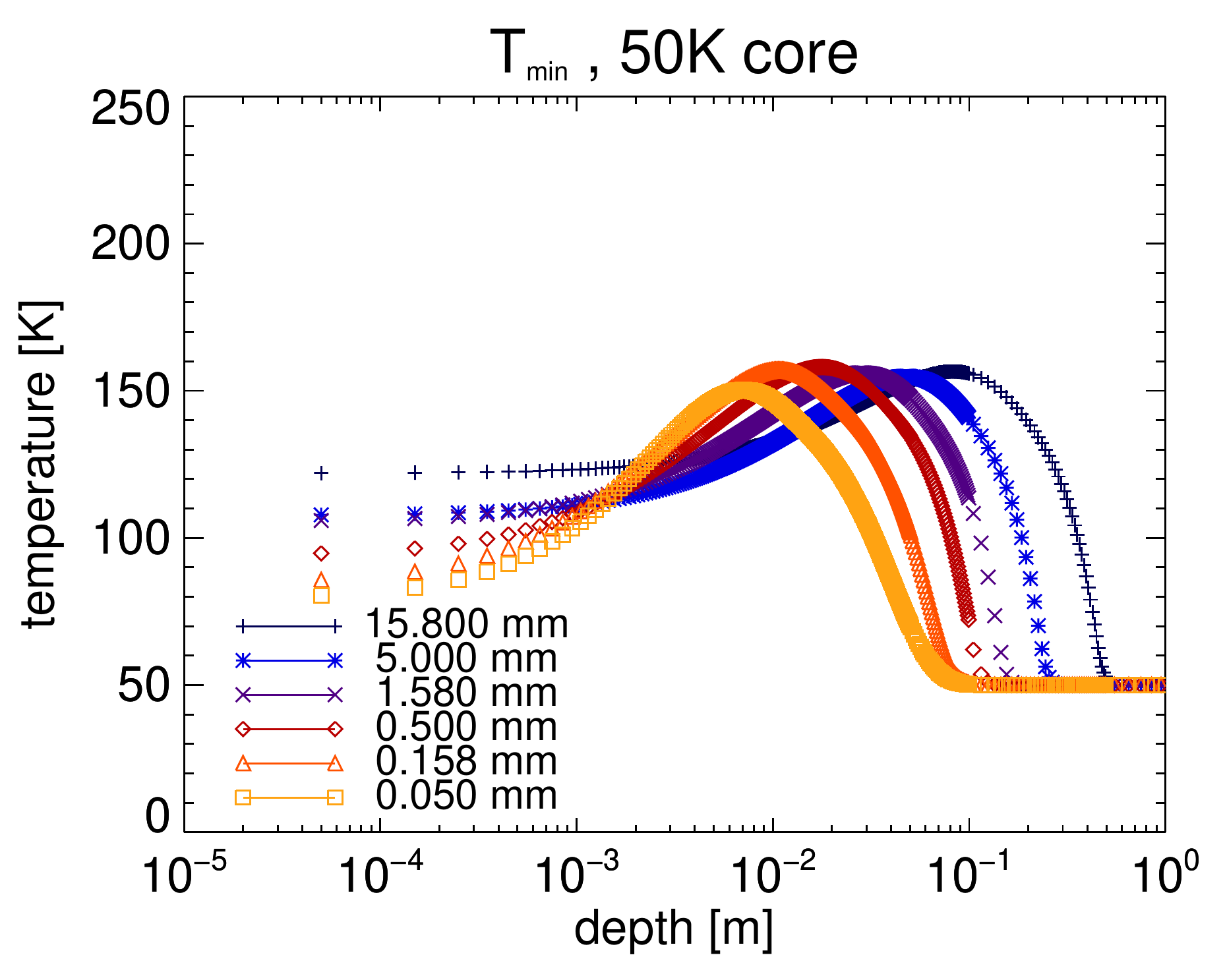}
    \includegraphics[width=8cm]{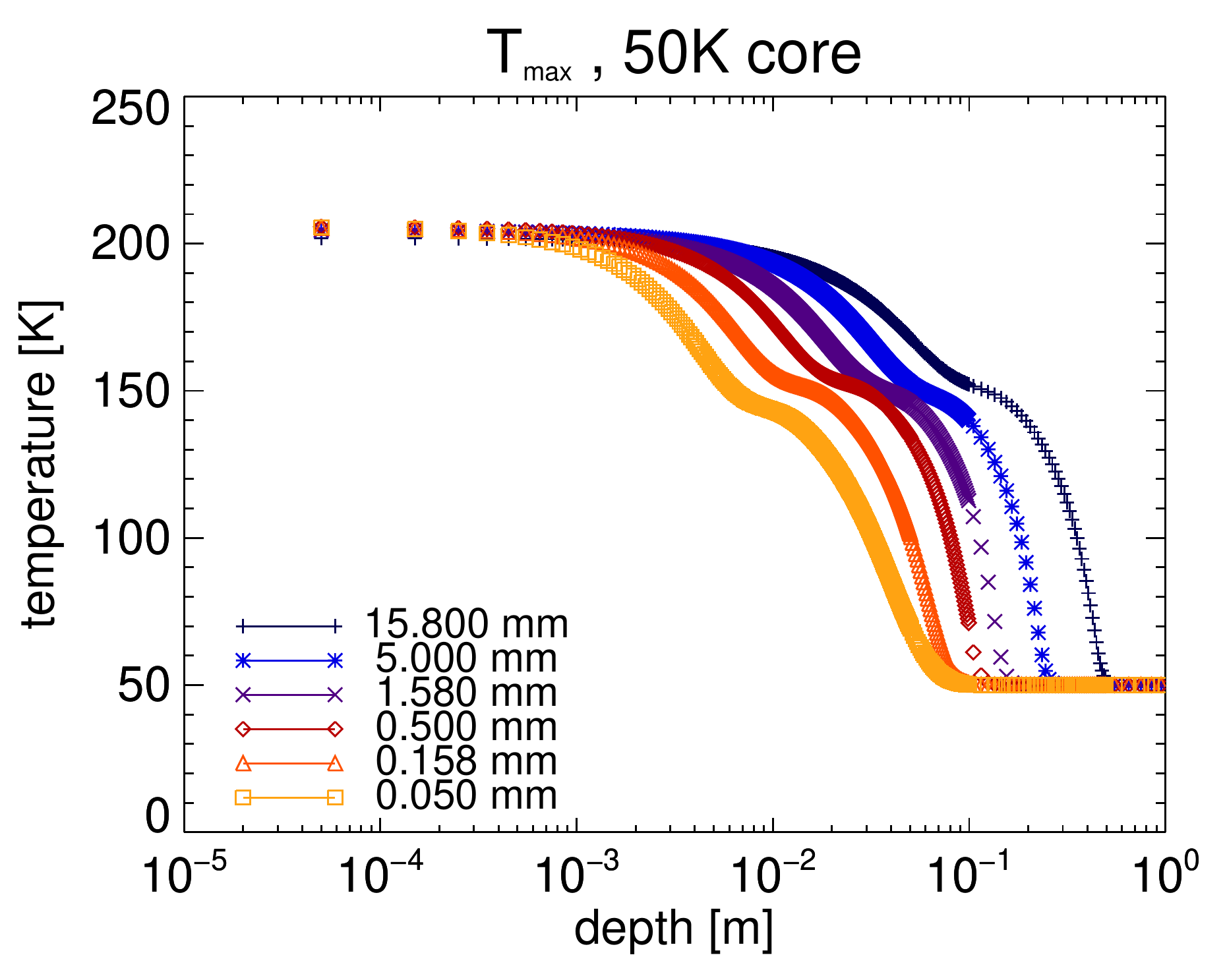}
	\includegraphics[width=8cm]{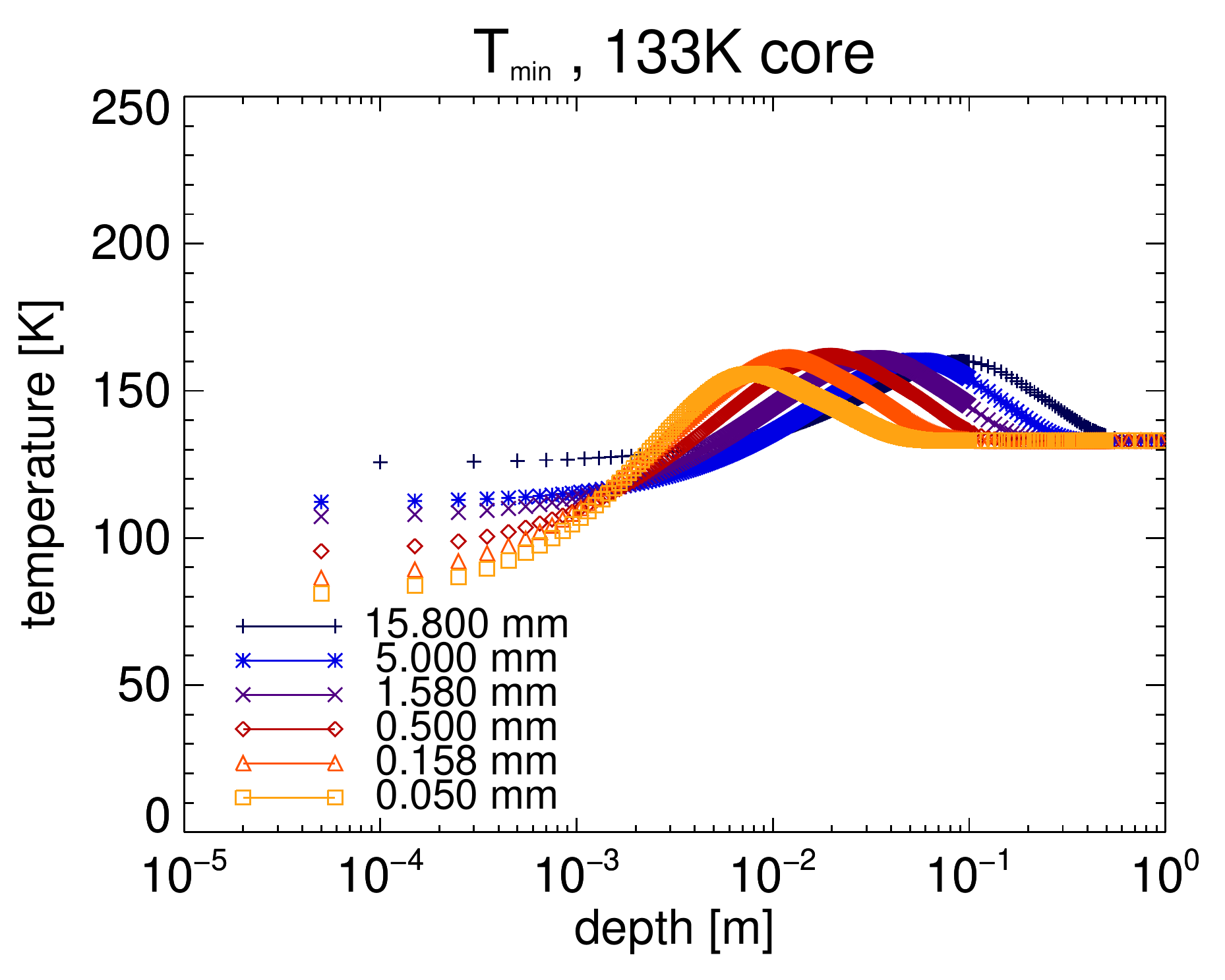}
    \includegraphics[width=8cm]{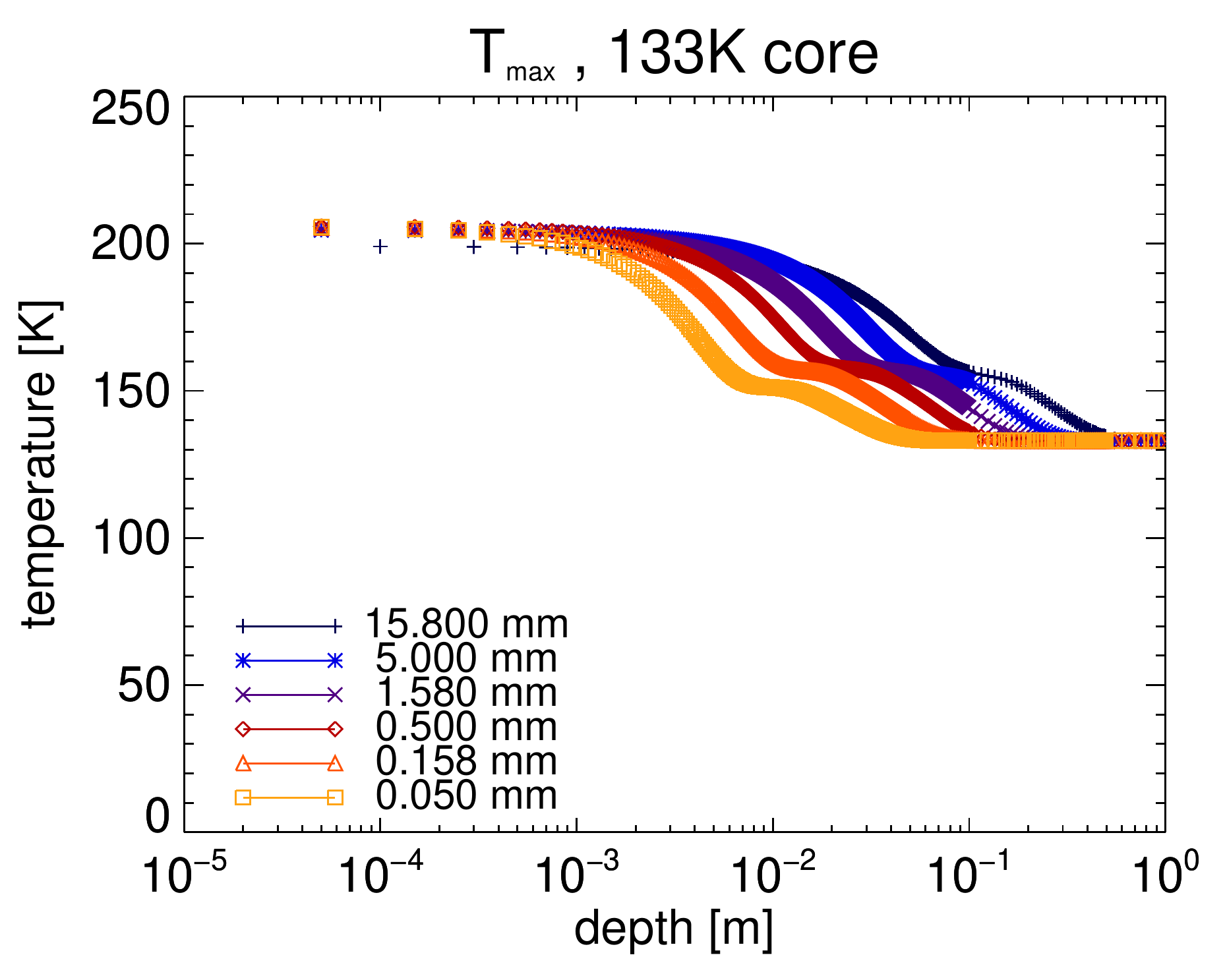}
    \caption{Snapshots of diurnal subsurface temperature profiles of a cometary nucleus that consists of dust aggregates of different sizes (marked by different colours, see legend) for minimum (left) and maximum (right) diurnal MIRO fluxes. The upper and lower plots show results for an initial temperature of 50 K and 133 K, respectively.}
    \label{fig:23}
\end{figure*}

The basic ingredients of our thermophysical model are: (1) Insolation. Depending on the position of the comet on its orbit, the location of the point considered on the nucleus surface, the local time at this location, and the Bond albedo, we determine the absorbed total solar flux (see left-hand side of Eq. \ref{Eq:02} in Appendix A). This incoming heat is balanced by direct re-emission from the comet surface (first term on the right-hand side of Eq. \ref{Eq:02}) and the amount of energy transported into the nucleus (second term on the right-hand side of Eq. \ref{Eq:02}). As the depth to which the diurnal and orbital heat waves penetrate into the comet interior is much smaller than the horizontal resolution of the MIRO instrument, we consider the heat transport a one-dimensional problem. (2) Heat transport mechanisms. The physical processes that we consider as most important to the (vertical) heat transport are conduction through the matrix of dust grains in contact (second term on the right-hand side of Eq. \ref{Eq:03}) and radiative heat exchange between the (assumed) dust pebbles (first term on the right-hand side of Eq. \ref{Eq:03}), respectively. We neglect heat flow by mass transport of gas and dust. Both terms that contribute to the heat conductivity (Eq. \ref{Eq:03}) are heavily dependent on the dust-aggregate size, which we assume to be monodisperse. Figure \ref{fig:07} shows this for three dust temperatures. For small aggregates, the heat conductivity decreases with increasing aggregate size, due to the decrease of the total inter-aggregate contact area per unit comet area (see dashed line in Figure \ref{fig:07}). For large dust aggregates, radiative heat exchange dominates over conduction, because it linearly depends on the size of the void spaces between aggregates, which naturally scale with aggregate size. However, radiative heat transport is temperature dependent as shown by the three curves in Figure \ref{fig:07}. The minimum heat conductivity is on the order of $10^{-4} ~ \mathrm{W~m^{-1}~K^{-1}}$ and is reached for aggregate radii of 0.1 mm and a temperature of 100 K. If no pebbles were present and the comet consisted of a homogeneous dust matrix, the heat conductivity would be $\sim 2 \times 10^{-3} ~ \mathrm{W~m^{-1}~K^{-1}}$ (horizontal solid line in Figure \ref{fig:07}). All material parameters that we used to derive these heat conductivities are shown in Table \ref{Table:01}. (3) Solving the heat-transport equation. Knowing the energy input and the basic heat-transport mechanisms, the heat-transport equation (Eq. \ref{Eq:09}) is solved using the Crank-Nicolson method (see Appendix A for details). This results in a vertical temperature stratification as a function of time for each dust-aggregate size (see Figure \ref{fig:23}). (4) Producing synthetic MIRO data. To compare the derived temperature profiles with the data acquired by the MIRO instrument \citep{gulkis2015}, the blackbody radiation emitted from different depths (and at different temperatures) inside the nucleus is integrated using the following emission-absorption algorithm,
\begin{equation} \label{Eq:01}
F_{\lambda}(T)=\int^{\infty}_0{{\alpha}_{\lambda}\ {B_{\lambda}\left(T\right)\ e}^{-{\alpha}_{\lambda}\ x}\ dx},
\end{equation}
where $x$ is the depth measured from the surface of the nucleus, ${\alpha}_{\lambda}$ is the length-absorption coefficient and $B_{\lambda}\left(T\right)$ is the Planck function for blackbody radiation. The integration is performed for the two different wavelengths used by the MIRO instrument, ${\lambda}_{\mathrm{sub-mm}}=\ 0.533\ \mathrm{mm}$ and ${\lambda}_{\mathrm{mm}}=\ 1.594\ \mathrm{mm}$, respectively.

The integration of Eq. \ref{Eq:01} is stopped at a finite depth when $F_{\lambda}(T)$ for the respective aggregate size saturates and, thus, no further flux contributions from deeper layers are expected.

We derived diurnal brightness temperatures by running the model until the diurnal heating and cooling cycle reached a steady state, with two initial temperatures of 50 K and 133 K. As we could not realistically model the long-term evolution of the internal temperature of the comet, we chose these two extreme cases, which represent a cold (50 K) and warm (133 K) storage of the nucleus of 67P in recent history. In Figure \ref{fig:23} we display two snapshots of the internal temperature distribution for each of the initial temperature conditions. The left two graphs show the time of minimum MIRO flux during a comet day at 3.27 au distance from the Sun, whereas the right two plots characterize the temperature stratification for the diurnal maximum MIRO flux. The coloured curves represent internal temperatures for different dust-aggregate sizes, as indicated in the legend of Figure \ref{fig:23}.

It turned out that the results presented below are rather insensitive to the actual choice of the internal starting temperature. As can be seen in Figure \ref{fig:23}, the temperature variation with depth at night (left two graphs in Figure \ref{fig:23}) is entirely the same for both initial conditions down to depths where the maximum temperature is achieved, which depends on the aggregate size, due to the size-dependent radiative term in the heat-transport equation (see Eqs. \ref{Eq:03} and \ref{eq:s2}). Beneath the temperature maximum, the temperature drops quickly with increasing depth to the initial value. In full sunlight (right two graphs in Figure \ref{fig:23}), temperatures systematically decrease from the surface downwards until they reach a plateau, which indicates the depth of the previous-day heat wave. Also here, there is no difference in temperature stratification between the two initial cases upward of this plateau. For deeper layers, the sudden drop to the initial temperature is similar to the night-time case. As most of the MIRO flux stems from the topmost few centimetres and increases with incrasing temperature, the actual choice of the initial condition makes only a small difference in the synthetic MIRO data (see Figure \ref{fig:04} below).

We then compared for both initial temperatures the expected monochromatic emission temperature of the comet for the two MIRO wavelengths, calculated following Eq. \ref{Eq:01}, with the subsurface temperature measurements by MIRO (see Figure 5 in \citet{gulkis2015}). We used a maximum illumination intensity of ${\mathrm{cos} \vartheta =0.96\ }$ in Eq. \ref{Eq:02}, representing the MIRO measurements \citep{gulkis2015} at northern latitude of $25^\circ$. As the criterion of match between the MIRO observations and model calculations, we chose the minimum and maximum diurnal temperatures measured by MIRO. These are for the sub-mm channel\footnote{The temperature values in the sub-mm channel were increased by 6\% relative to \citet{gulkis2015}, due to a 94\% beam efficiency, see \citet{schloerb2015}. $T_{\mathrm{min}}=(122 \pm 3) \ \mathrm{K}$ and $T_{\mathrm{max}}=(184 \pm 3)\ \mathrm{K}$ and for the mm-channel $T_{\mathrm{min}}=(140 \pm 3)\ \mathrm{K}$ and $T_{\mathrm{max}}=(162 \pm 3) \ \mathrm{K}$}, respectively. We systematically varied the absorption coefficient for the material in the range ${\alpha }_{\mathrm{sub-mm}}=0.25\ {\mathrm{cm}}^{-1}$ to $50\ {\mathrm{cm}}^{-1}$ (in steps of $0.25\ {\mathrm{cm}}^{-1}$) for the sub-mm channel and ${\alpha }_{\mathrm{mm}}=0.10\ {\mathrm{cm}}^{-1}$ to $50\ {\mathrm{cm}}^{-1}$ (in steps of $0.10\ {\mathrm{cm}}^{-1}$ for $\alpha_{\mathrm{mm}} \le 1.0\ \mathrm{cm}^{-1}$ and in steps of $1.0\ \mathrm{cm}^{-1}$ for $\alpha_{\mathrm{mm}} > 1.0\ \mathrm{cm}^{-1}$) in the mm channel. Pebble radii were varied between $a=50\ \mathrm{\mu m}$ and $a=1.58\ \mathrm{cm}$ in logarithmically equidistant steps of $\sqrt{10}$.

\begin{figure*}
	\includegraphics[width=\columnwidth]{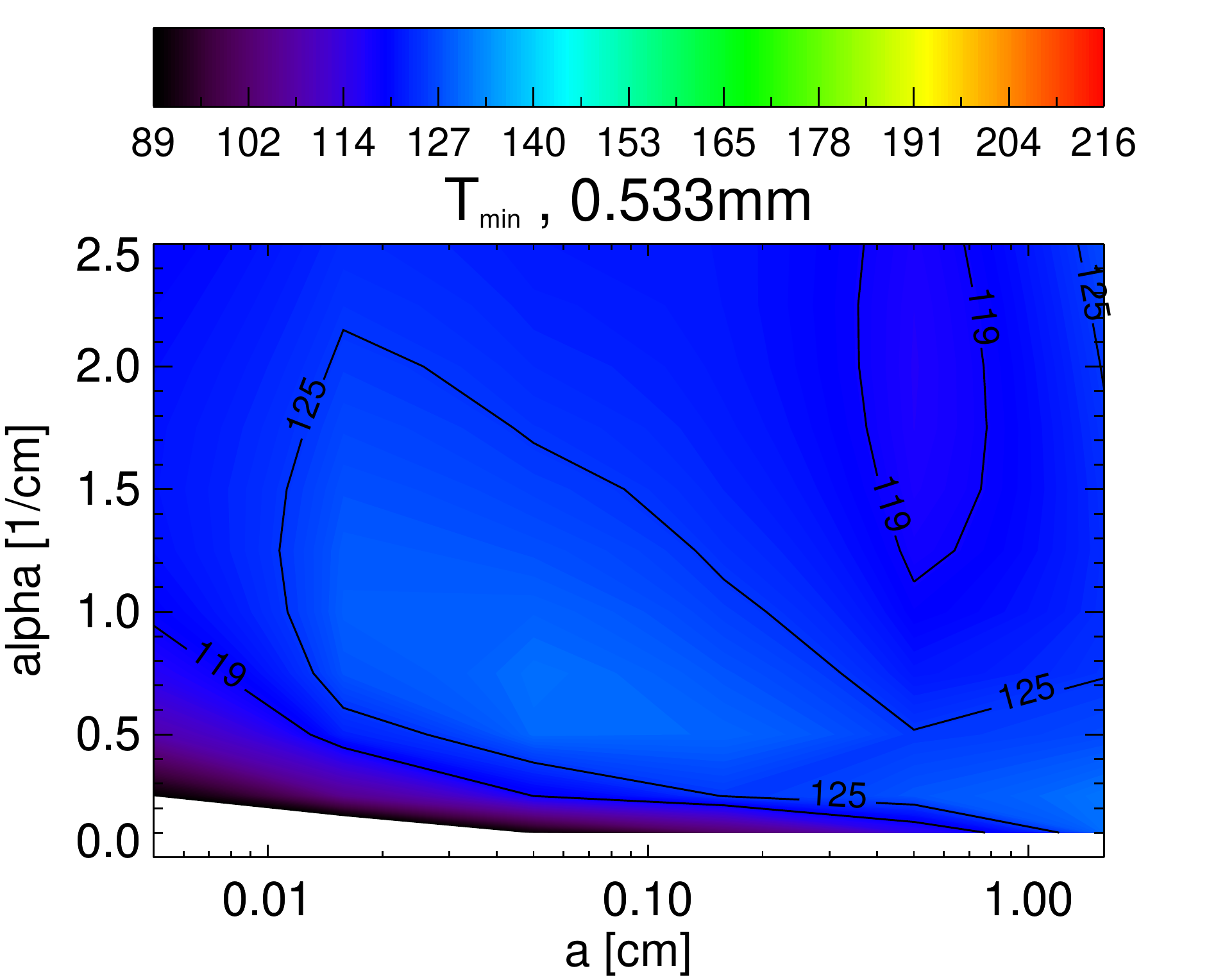}
    \includegraphics[width=\columnwidth]{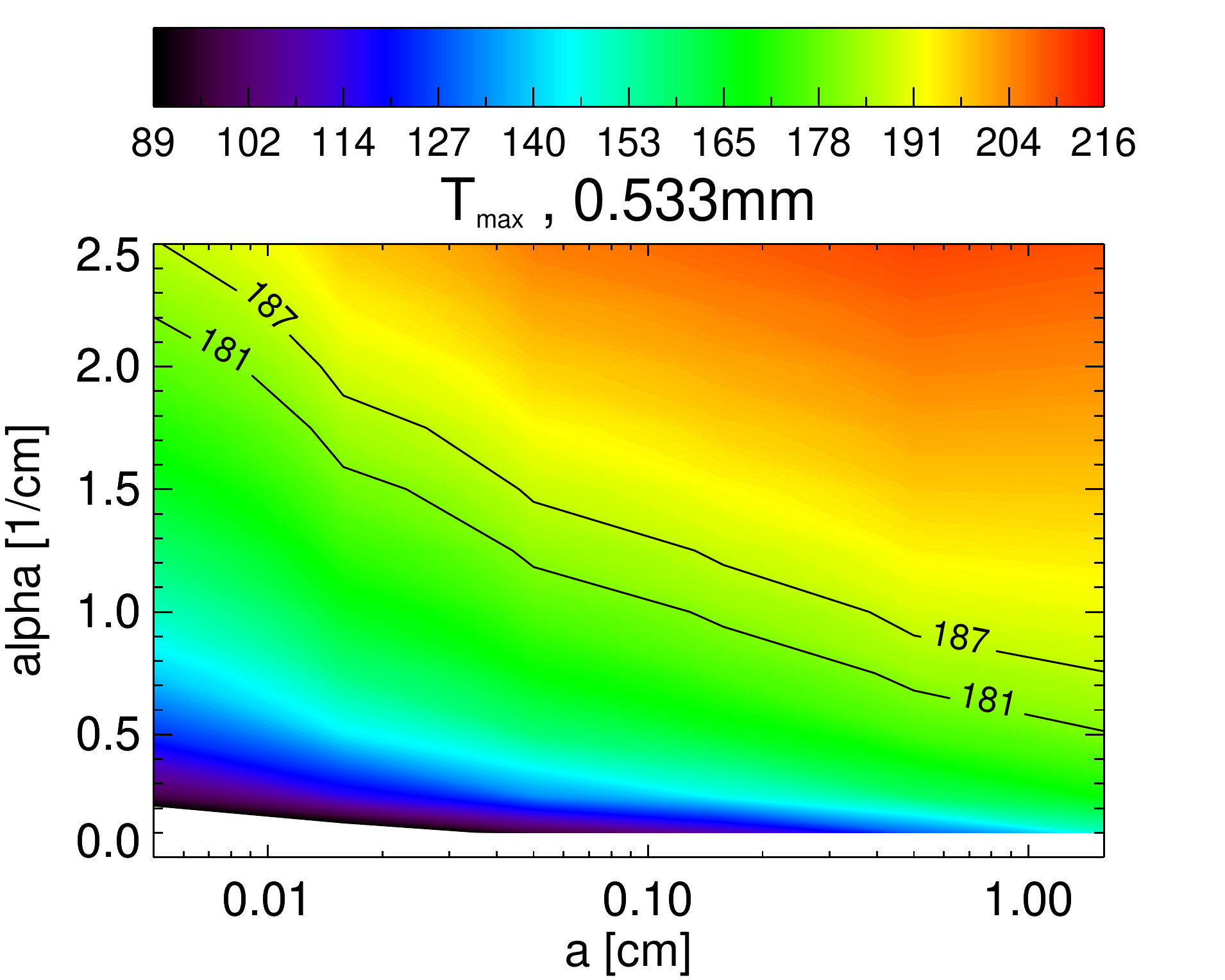}
   	\includegraphics[width=\columnwidth]{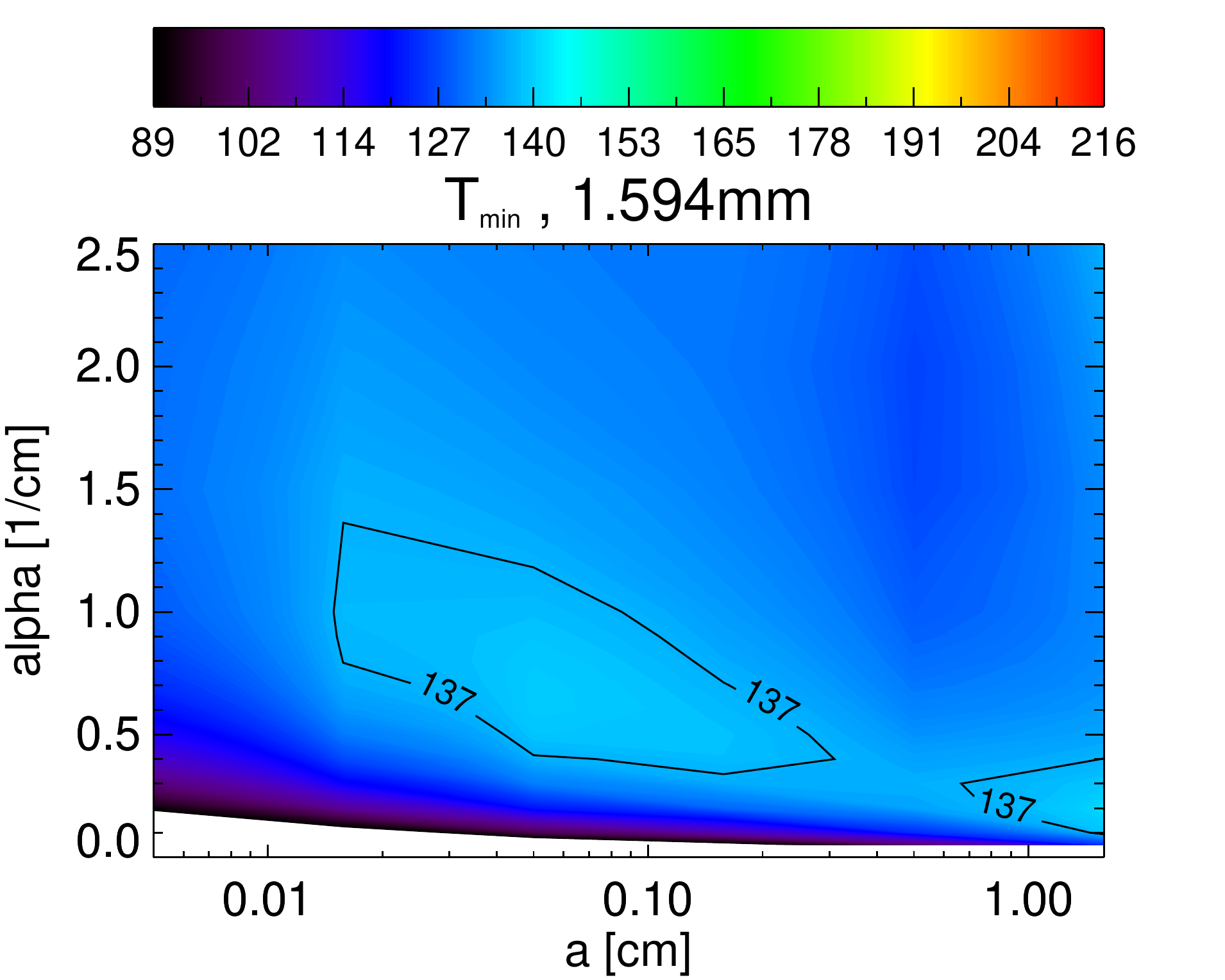}
    \includegraphics[width=\columnwidth]{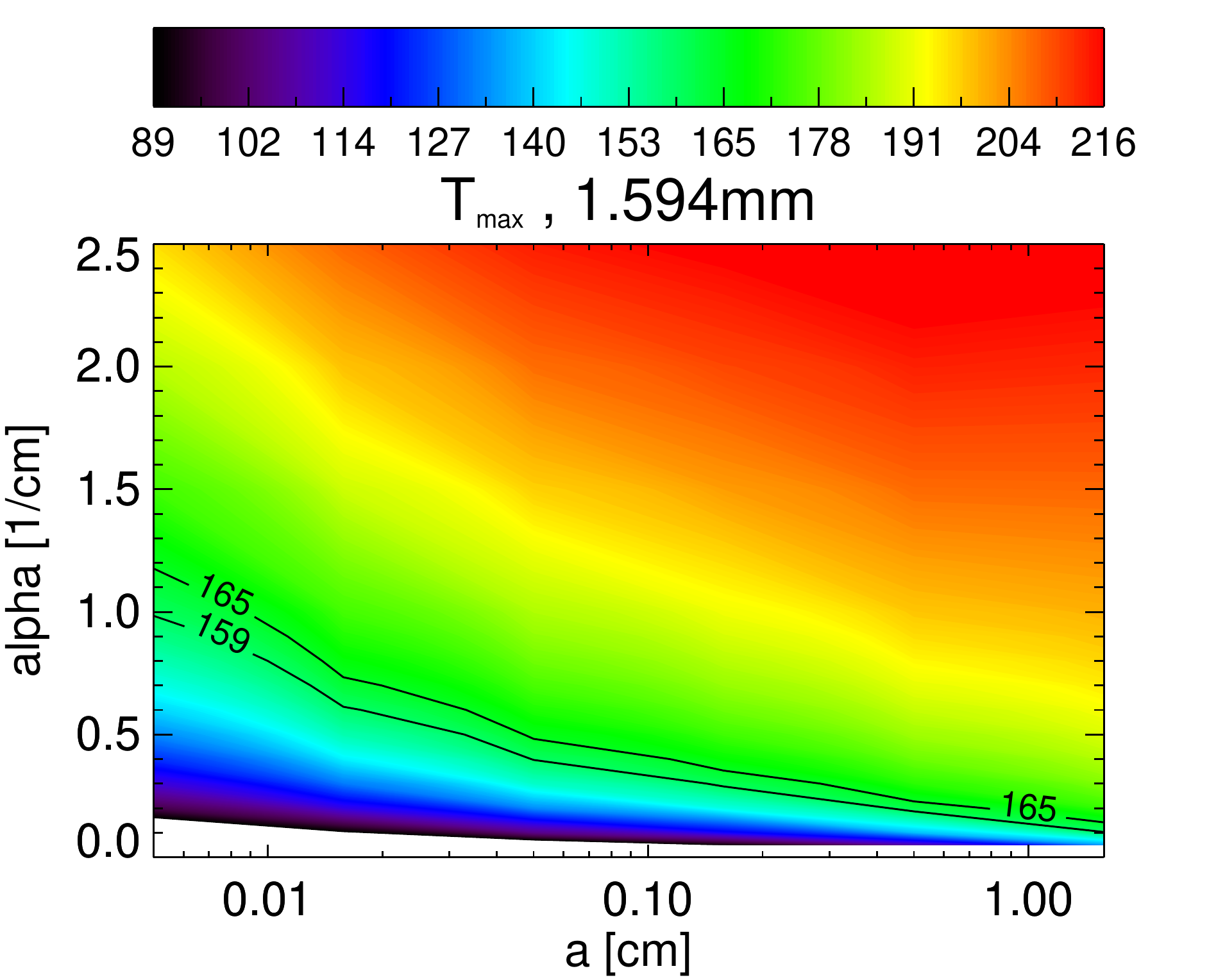}
    \caption{Simulation results for minimum and maximum diurnal MIRO temperatures for an initial temperature of 50 K. Minimum (left two panels) and maximum diurnal temperatures (right two panels) are indicated by colours (see colour bar) as a function of pebble radius (x axis) and absorption coefficient (y axis) for the sub-mm channel (upper two panels) and the mm-channel of MIRO (lower two panels). The contours mark the two extreme temperatures that are in agreement with the MIRO data, i.e., a good match between synthetic and measured data can be found between the contours.}
    \label{fig:03}
\end{figure*}

Figure \ref{fig:03} shows a mosaic of derived minimum and maximum temperatures for the two MIRO wavelengths in the aggregate-size and absorption-coefficient range described above and for an initial temperature of 50 K. For both MIRO channels, an anti-correlation between the required absorption coefficient and the aggregate size can be discerned. This degeneracy is evident when Figure \ref{fig:23} is examined. Due the dominance of radiative heat transport and the corresponding linear dependency of the heat conductivity on the pebble size (see Appendix A), the diurnal penetration depth of the heat wave scales with the square root of the pebble radius. To achieve the same synthetic MIRO temperature, the absorption coefficient therefore has to decrease with increasing dust-aggregate size. The two contours in each of the four panels of Figure \ref{fig:03} denote the boundaries of the measured MIRO temperature ranges as defined above. A formal agreement between the synthetic and measured MIRO data can be achieved in both MIRO channels for dust-aggregate sizes of a few millimetres. However, while the range of maximum synthetic MIRO temperatures is quite large for the various aggregate sizes and absorption lengths, the minimum temperatures are relatively insensitive to changes in these parameters (see Figure \ref{fig:03}). Thus, the MIRO minimum-temperature measurements do not efficiently constrain the aggregate sizes so that we need additional aggregate-size restrictions, which will be discussed hereafter.

A comparison of the two MIRO channels for the maximum MIRO temperatures is shown in Figure \ref{fig:04}, where the red (sub-mm channel) and orange (mm channel) bands in the plot denote the aggregate-size and absorption-coefficient ranges for which an agreement between the MIRO measurements and the simulations could be achieved for both starting temperatures. As noted above, we cannot derive the aggregate sizes on comet 67P with this data alone, due to the degeneracy between aggregate size and absorption length. However, we can use additional information on the absorption coefficient to exclude very large and very small aggregate sizes, respectively. Assuming a dust-aggregate radius $a$ and optically thick aggregates, the maximum absorption coefficient is given by geometrical optics and reads $\alpha_{\mathrm{max}}= n ~ \pi ~ a^2$, with $n$ being the number density of the aggregates. We assume a packing density of $\phi=0.6$, which is close to the random close packing limit (see Sect. \ref{Sect:porosity}), and get $\alpha_{\mathrm{max}}= \frac{3~ \phi}{4~a} = \frac{0.45}{a}$. This relation between the absorption coefficient and aggregate size is shown as a solid black line in Figure \ref{fig:04}. It is evident that for aggregate sizes larger than about $a \approx 6 \ \mathrm{mm}$ (independent of the initial temperature), the calculated absorption coefficient exceeds the geometric value for perfectly absorbing particles so that we can exclude aggregates with these sizes. In the above analysis, we neglected surface reflection, which in principle can enlarge the distance travelled by the heat wave before it escapes the surface over the geometrical depth. However, this effect should be small for mm-sized aggregates and geometrical depths of $\sim$ 10 cm, because of the rather low permittivity of the nucleus material. For $\epsilon = 1.27$, as measured by CONSERT \citep{kofman2015}, and under the assumption that this value does not change too much from metre to millimetre wavelengths, we get a surface reflectivity of only 0.4\%, which leads to a total reflection loss over 10 cm depths of $\sim 8\%$ for aggregates with $a=5 \ \mathrm{mm}$.

The ratio between the calculated values of the absorption coefficient for the two MIRO wavelengths, $\alpha_\mathrm{sub-mm}/\alpha_\mathrm{mm}$ can be estimated from Figure \ref{fig:04}. The blue, turquoise and green curves denote $2.2 \times$, $3.2 \times$ and $4.2 \times \alpha_{\mathrm{mm}}$ (50 K initial temperature, upper plot) and $2.9 \times$, $4.3 \times$ and $5.8 \times \alpha_{\mathrm{mm}}$ (133 K initial temperature, lower plot) the mm-curve, respectively. We can derive from Figure \ref{fig:04} that $\alpha_\mathrm{sub-mm}/\alpha_\mathrm{mm} \approx 2.2 \ldots 4.2$ ($T=50$ K) and $\alpha_\mathrm{sub-mm}/\alpha_\mathrm{mm} \approx 2.9 \ldots 5.8$ ($T=133$ K), respectively. As the ratio of the MIRO frequencies $\nu$ is exactly 3, we get a spectral index of $\beta' = \frac{\mathrm{d}\log\alpha}{\mathrm{d}\log\nu} = 0.7 \ldots 1.3$ ($T=50$ K) and $\beta' = \frac{\mathrm{d}\log\alpha}{\mathrm{d}\log\nu} = 1.0 \ldots 1.6$ ($T=133$ K), with the lower $\beta'$ values for the smaller dust aggregates.

\citet{draine2006}, \citet{natta2007} and \citet{testi2014} showed that a value of $\beta' \approx 1$ for wavelengths around 1 mm is characteristic for MRN-type size distributions with size cut-off of $a_{\mathrm{max}} \gtrsim 1$ mm. For size distributions with smaller maximum sizes, $\beta'$ typically reaches values of $\beta' \approx 2$ for mm wavelengths. \citet{menu2014} showed that the resulting dust opacities (see below) are in agreement with models of mixtures of amorphous silicates and carbonaceous material if the contributions from sub-mm-sized dust aggregates are negligible. From this analysis, we conclude that the most likely aggregate sizes to explain the MIRO measurements fall into the range $a \approx 1$ mm to $a \approx 6$ mm. We should like to point out that in the above analysis, we ignored electromagnetic interference between neighbouring dust aggregates. Although this might be a crude approximation, near-field effects are probably of small importance for materials with small index of refraction, as suggested by the low opacities at mm wavelength and low permittivity of $\epsilon = 1.27$ measured by CONSERT at 3.3 m wavelength \citep{kofman2015}.

It should also be mentioned that for this analysis, we ignored the heat sink due to ice sublimation because of the low outgassing rate at 3.27 au. Details about the influence of outgassing rate on the internal temperatures can be found in Appendix A. However, future investigations will have to take the latent heat of water ice and the ice-dust boundary as a free parameter into account.

With a mass density of comet 67P of $0.53\ \mathrm{g\ }{\mathrm{cm}}^{-3}$ \citep{jorda2016}, we convert the derived absorption coefficients $\alpha_{\mathrm{sub-mm}} \approx 0.8-1.2~\mathrm{cm^{-1}}$  and $\alpha_{\mathrm{mm}} \approx 0.2-0.4~\mathrm{cm^{-1}}$ for a dust-aggregate size range of $a=1-6$ mm (see Figure \ref{fig:04}) into mass-absorption coefficients (opacities) of $\kappa_{\mathrm{sub-mm}} \approx 1.5-2.3\ {\mathrm{cm}}^2\ {\mathrm{g}}^{-1}$ and $\kappa_{\mathrm{mm}}=0.4-0.8\ {\mathrm{cm}}^2\ {\mathrm{g}}^{-1}$, respectively. These values are slightly higher than those calculated by \citet{draine2006} for astrosilicate, but seem to be typical for carbonaceous material, and provide for the first time direct estimates of these important properties (albeit for a complex hierarchical arrangement of aggregates), which will be helpful in future modelling of dust processes in protoplanetary and debris discs. As mentioned above, the opacities derived by \citet{menu2014} for a mixture of silicates and carbonaceous material are in broad agreement with our estimates when the aggregates are typically millimetre-sized.

\begin{figure}
	\includegraphics[width=\columnwidth]{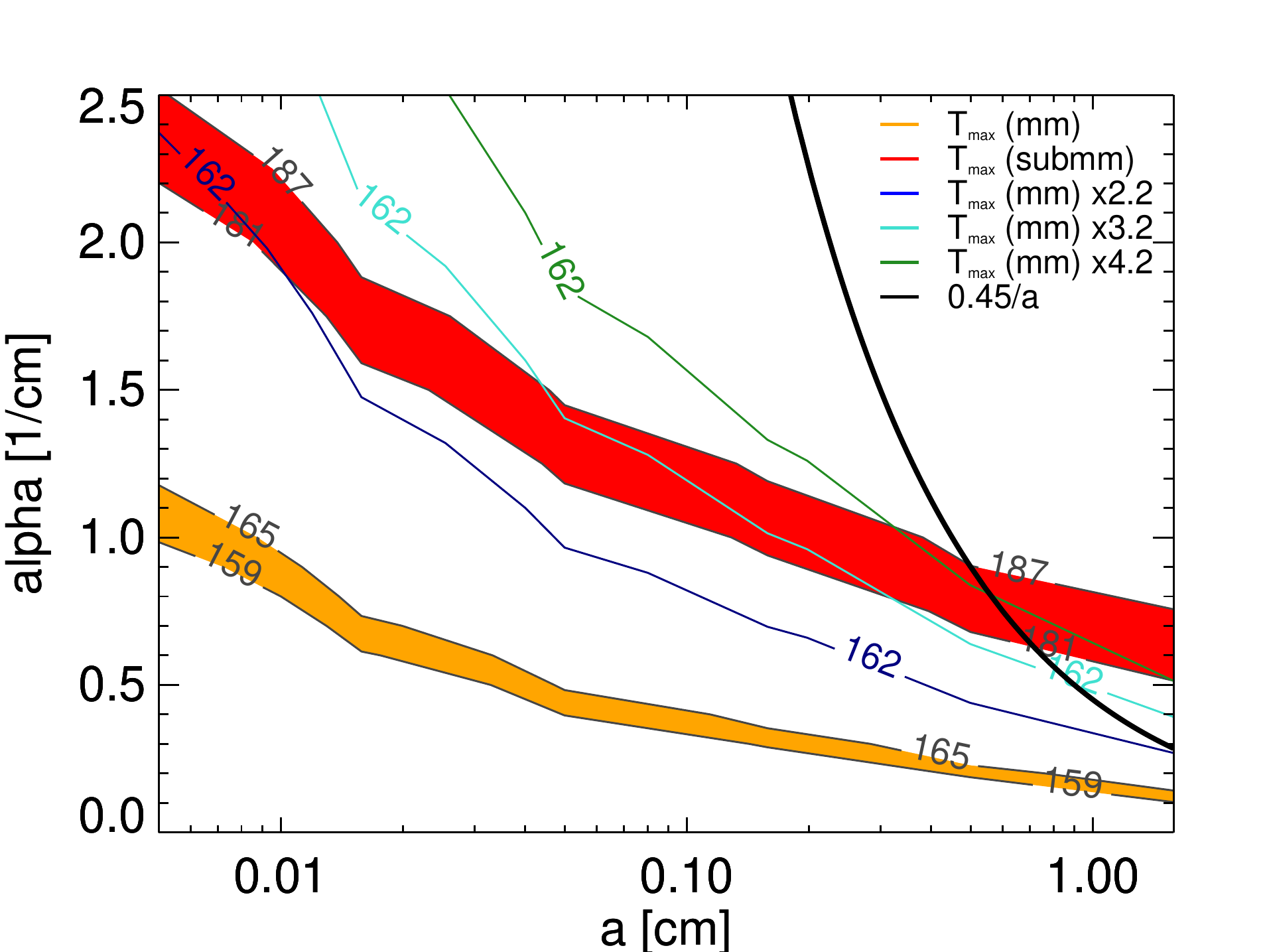}
    \includegraphics[width=\columnwidth]{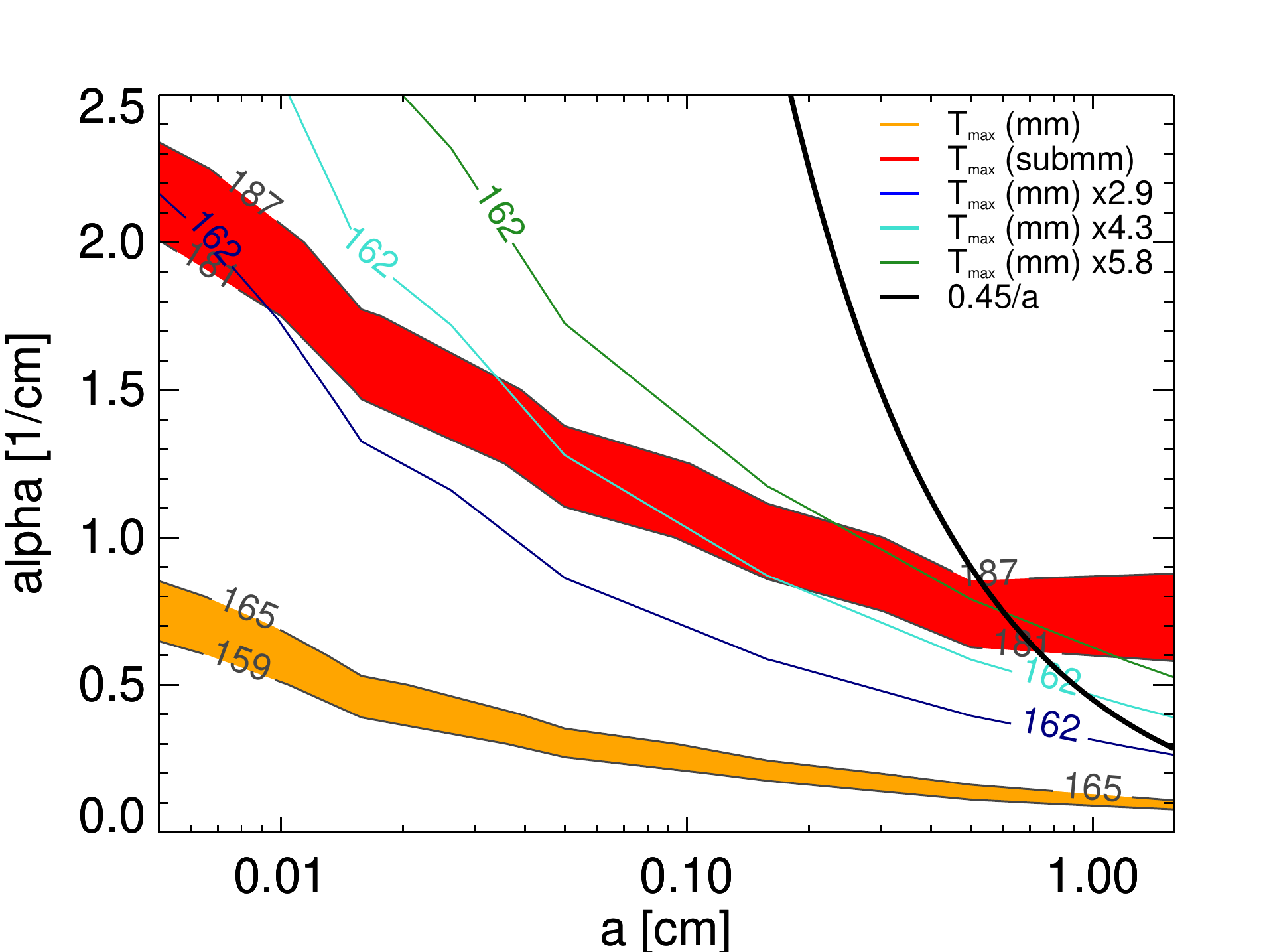}
    \caption{Comparison between measured and modelled maximum diurnal MIRO temperatures. As broad coloured bands, pebble radii (x axis) and absorption coefficients (y axis) are shown for which the synthetic and measured maximum temperatures overlap for the sub-mm channel (upper red band) and the mm-channel (lower orange band) of MIRO. The blue, turquoise and green curves denote $2.2 \times$, $3.2 \times$ and $4.2 \times \alpha_{\mathrm{mm}}$ (50 K initial temperature, upper plot) and $2.9 \times$, $4.3 \times$ and $5.8 \times \alpha_{\mathrm{mm}}$ (133 K initial temperature, lower plot) the mm-curve, respectively. The black curve denotes the maximum possible absorption length (see text for details).}
    \label{fig:04}
\end{figure}

\textit{4. Pebble-size constraint from the surface heating curve measured by Philae.} The infrared radiometer MUPUS-TM recorded the surface temperature of the landing site of Philae \citep{spohn2015} over a total period of 40 h. Although the observed surface was most of the time in shadow, for short ($\sim$40 min) periods of insolation, the temperature rose from $\sim 117\ \mathrm{K}$ to $\sim 129\ \mathrm{K}$ and then dropped off again. As MUPUS-TM recorded thermal-infrared radiation, we modelled by the finite element method (FEM) the temperature of a surface layer of equal-sized spherical pebbles when the insolation is momentarily switched on (see Appendix B).

Figure \ref{fig:05} shows the values of the mean squared temperature deviation between the measurement and the FEM model as a function of pebble radius for the case of full illumination of the area observed by MUPUS-TM ($f_{\mathrm{s}}=1$ in Eq. \ref{Eq:17}). The best-fitting value of $a\ =\ 0.44\ \mathrm{mm}$ is clearly the global minimum within four decades of realistic radii. However, a wide range of pebble radii from $a\ =\ 0.22\ \mathrm{mm}$ to $a\ =\ 55\ \mathrm{mm}$ solve the equations with $\left\langle {\Delta T}^2\right\rangle \lesssim 1\ {\mathrm{K}}^2$, which is about the noise level in the MUPUS-TM data. Outside this radius range, the mean squared temperature deviation increases drastically (see Figure \ref{fig:05}). Smaller dust aggregates heat too fast to fit the measurements, whereas for larger pebbles the heat wave does not penetrate deep enough, which also leads to a mismatch between the measured and modelled temperatures.

\begin{figure}
	\includegraphics[width=8cm]{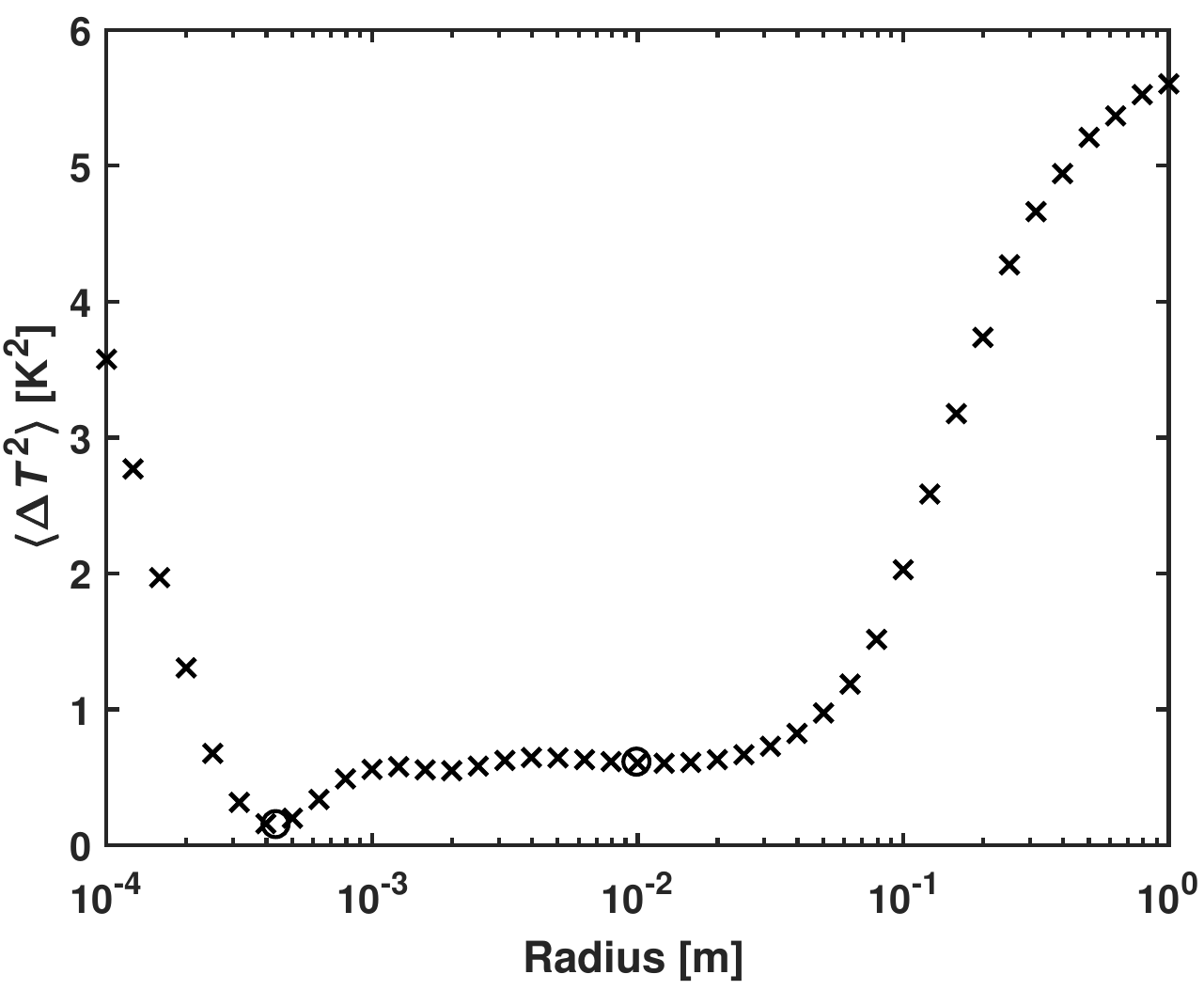}
    \includegraphics[width=4cm]{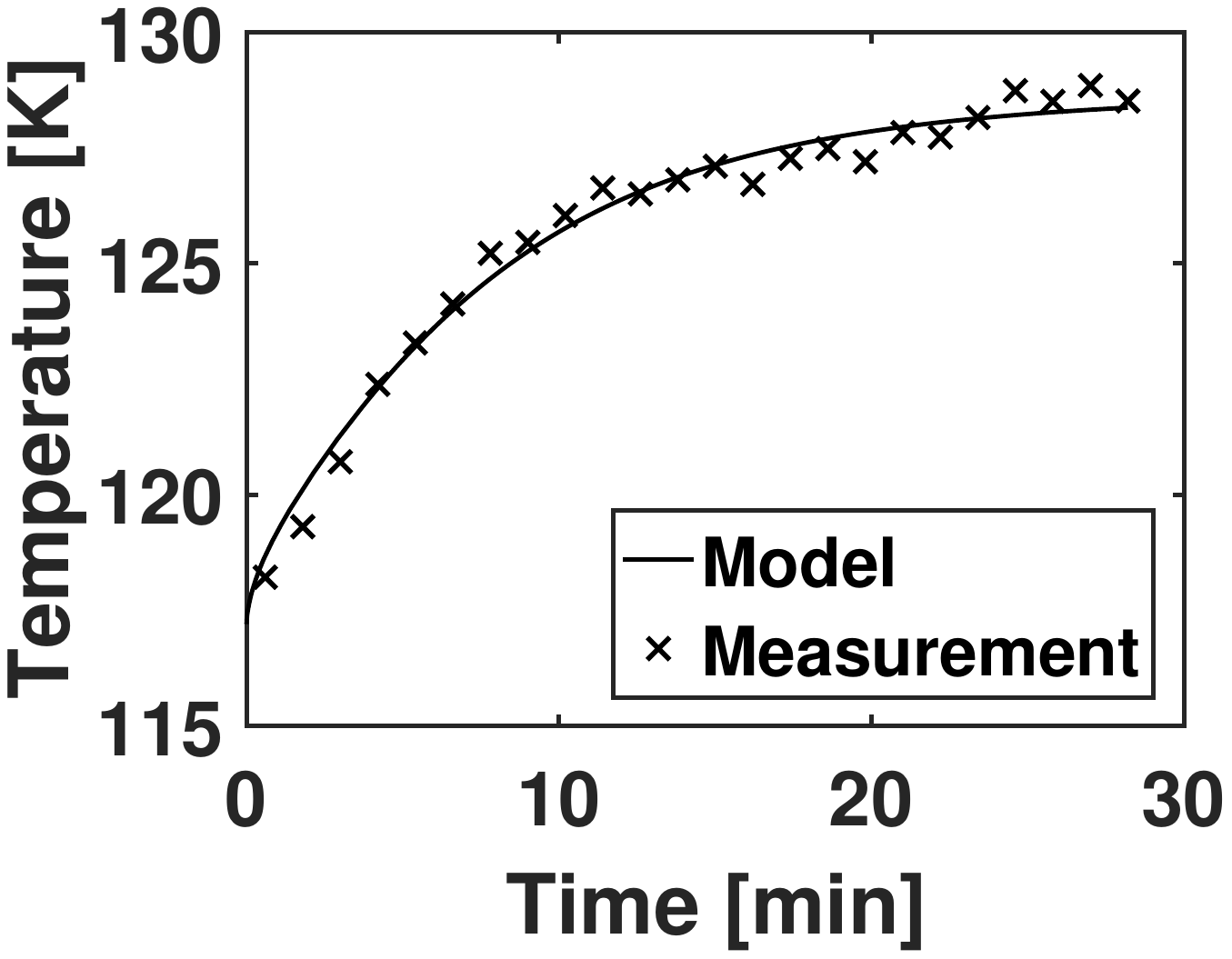}
    \includegraphics[width=4cm]{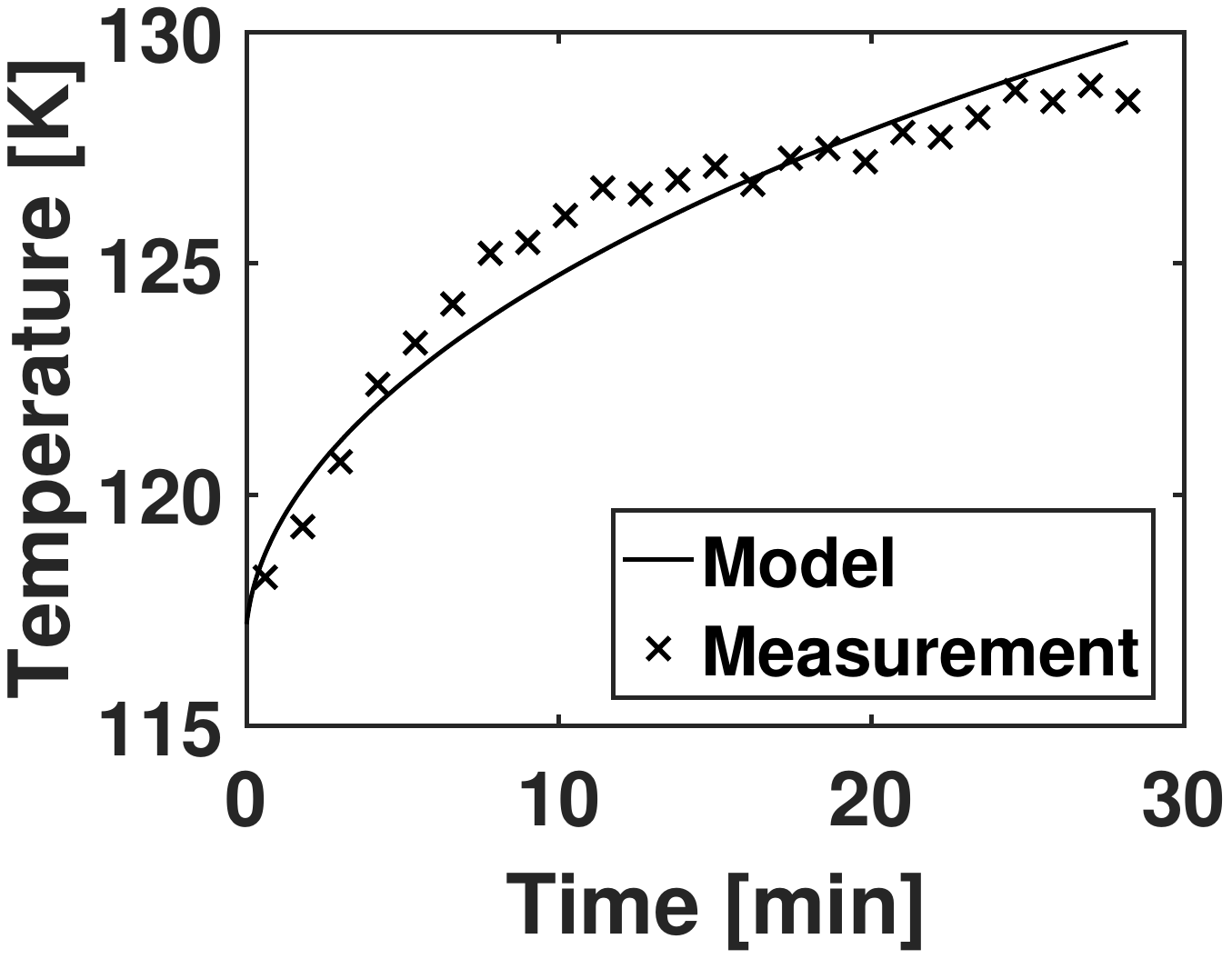}
    \caption{Comparison between MUPUS-TM data and the model. Top: Mean squared temperature deviation between the measurements and the FEM model of the comet surface for different pebble radii. Bottom: Surface temperature of comet 67P measured by MUPUS-TM (crosses, \citet{spohn2015}) and best-fitting model (solid curves) with a pebble radius of $a\ =\ 0.44\ \mathrm{mm}$ (bottom left) and $a\ =\ 10\ \mathrm{mm}$ (bottom right).}
    \label{fig:05}
\end{figure}

The goodness of fit can also be seen in Figure \ref{fig:05}, where the MUPUS-TM data from \citet{spohn2015} are shown as crosses together with the results of the FEM model (solid curves) for the best-fitting radius of $a\ =\ 0.44\ \mathrm{mm}$ and an illumination factor of $f_{I_{\odot }}$= 0.16 (see Appendix B) as well as for a pebble with a radius of $a=10\ \mathrm{mm}$ and an illumination factor of $f_{I_{\odot }}$=0.24. The same values were achieved for both limiting cases of the ambient temperature used for the heat exchange of the bottom sphere surface discussed in item (iii) of the model assumptions (see Appendix B). Deviations from the illumination of the whole observation area of the MUPUS-TM sensor, as described by Eq. \ref{Eq:17}, with either a constant or a time-dependent fractional illuminated area do not influence the result for the size distribution of pebbles.

If we assume that the comet-nucleus surface does not consist of spherical pebbles, but is made of a layer of $\rm \mu m$-sized dust particles with a filling factor of $\phi = 0.4$, we can use the model shown in Appendix B and derive the optimal thickness of this dust layer to match the temperatures observed by MUPUS-TM. We get a reasonable fit (with a $\left\langle {\Delta T}^2\right\rangle$ very similar to that shown in Figure \ref{fig:05}) when the half-thickness of the dust layer is between 100 $\rm \mu$m and 10 cm. For a thickness outside this range, $\left\langle {\Delta T}^2\right\rangle$ increases similarly rapidly as in the case with pebbles. Thus, the measurements by MUPUS-TM indicate a characteristic length scale as shown in Figure \ref{fig:02}.

\begin{figure}
	\includegraphics[width=\columnwidth]{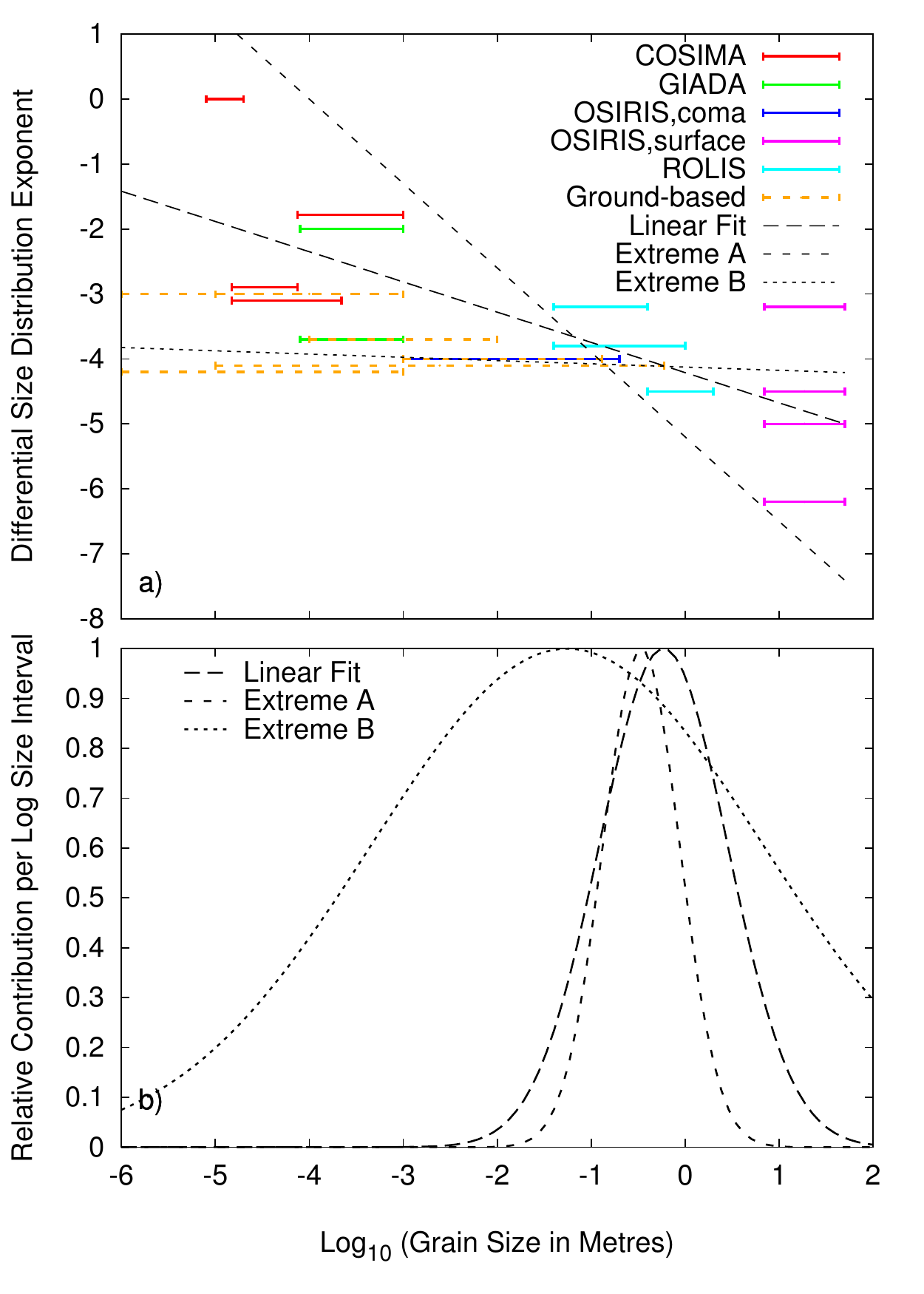}
    \caption{Particle size distribution for comet 67P. (a) Exponent of the size-frequency distribution function, $\beta$, of the dust emitted from the nucleus of comet 67P. Data (horizontal lines) stem from measurements by various instruments on board Rosetta and from Earth-based observations. The three lines represent a linear fit to the data and two extreme approximations, one very steep and one very shallow. (b) Derived normalized mass-frequency distributions per logarithmic size interval for the three linear approximations shown in panel (a). For a flat line at $\beta$ = 4 in panel (a), all size bins would contribute equally to the mass.}
    \label{fig:09}
\end{figure}

\textit{5. Pebble-size constraint from the size distribution of dust observed by Rosetta.} The various dust-sensitive instruments on-board Rosetta and Philae (COSIMA, GIADA, OSIRIS, ROLIS) have detected individual size-frequency distributions of the dust over many orders of magnitude in diameter (see Appendix C). Taking these individual data sets and fitting piecewise power laws of the form $n\left(a\right)\mathrm{d}a \propto a^{\beta} \mathrm{d}a$, with $n(a)\mathrm{d}a$ being the number of detected dust particles in the radius range between $a$ and $a+\mathrm{d}a$, shows that the exponent of the size-frequency distribution, $\beta$, systematically varies with the dust size (see discussion in Appendix C). A power-law size-frequency distribution translates into a total mass per logarithmic size bin of $M(a)\propto n\left(a\right)a^4\propto a^{4+\beta}$. For $\beta < -4$, the logarithmic mass distribution function is dominated by the small-mass end, whereas for $\beta > -4$, the large particles dominate. The mass-dominating grains with $\beta \approx -4$ are present in the size range between millimetres and a few metres (see Figure \ref{fig:09}). Although one has to be careful with the assumption of a single continuous mass-distribution function throughout the observable size range, it is clear that pebble-sized dust aggregates are very abundant in comet 67P.

\textit{6. Pebble-size constraint from measurements of the tensile strength of comet 67P.} For small bodies, it is not only gravity that keeps the objects in shape, but the internal cohesion plays an important role \citep{gundlach2015b}, which is caused by the mutual van der Waals attraction between dust grains in contact. A measure of the cohesion is the tensile strength of the material, i.e., the detachment force per effective particle cross section. The various Rosetta instruments derived tensile strengths of $3-15\ \mathrm{Pa}$ and $10-20\ \mathrm{Pa}$ for the 67P surface material on length scales of 10s of metres \citep{groussin2015,thomas2015} and $10-200\ \mathrm{Pa}$ for the bulk comet \citep{hirabayashi2016}.

In general, the tensile strength of a particulate material is determined by the number of grain-grain contacts per unit cross section. For a homogeneous material with no pebbles, only the packing density of the grains can alter the bulk cohesion. As the typical cohesion force between $\rm \mu$m-sized silica grains is on the order of ${10}^{-7}\ \mathrm{N}$ \citep{heim1999}, a network of these grains with a filling factor of $\phi=0.20-0.54$ possesses a tensile strength of $1000-3700\ \mathrm{Pa}$ \citep{blum2004}. This was also recently confirmed to be the case for the sub-mm-sized dust aggregates entering the COSIMA instrument of Rosetta \citep{hornung2016}. Organic material may slightly lower these values, whereas for water ice in hexagonal form and for temperatures $\lesssim 200$ K, relevant for the sub-surface activity regions of comet 67P, tensile strengths will be a factor $\sim$10 higher \citep{gundlach2011,gundlach2015}. Monomer-grain sizes smaller than $\sim$1 $\rm \mu$m will also increase the tensile strength.

However, if the comet nucleus consists of a hierarchical arrangement of pebble-sized dust aggregates, it is the cohesion between these aggregates that determines the overall tensile strength of the comet, although the pebbles themselves possess much higher internal strengths \citep{blum2014,blum2015}. For low compression (as is the case for cometary nuclei, see Sect. \ref{Sect:homogeneity}), the contact area between two touching aggregates is much smaller than their cross section so that the bulk tensile strength is reduced in comparison to a homogeneous dusty body. For spherical dust aggregates of radius $a$ consisting of $\rm \mu$m-sized silica particles, the bulk tensile strength of a cluster of pebbles with $\phi_{\mathrm{pebble}}=0.4$ and packed with $\phi=0.6$ is
\begin{equation}
\label{Eq:tensilestrength}
\Sigma=0.21\ \mathrm{Pa\ }\times \ {\left(\frac{a}{1\ \mathrm{cm}}\right)}^{-2/3}
\end{equation}
\citep{skorov2012,blum2014}.

Thus, the derived tensile-strength values for comet 67P (see above) cannot be reconciled with a homogeneous dusty composition of the nucleus. This rules out any collisional-growth-only models for planetesimals \citep[see, e.g.,][]{windmark2012}, which rely on mass accretion in high-velocity impacts, because they compress the material to its densest state of $\phi \approx 0.4$ \citep{guettler2010} and lead to too high tensile strengths and bulk porosities much lower than the measured values for
comet 67P \citep{fulle2016a,kofman2015,paetzold2016}.

Taking into account that the derived bulk value for the tensile strength of comet 67P \citep[see][]{hirabayashi2016} is also influenced by gravity -- the central lithostatic compression of the large lobe of comet 67P is $\sim$86 Pa (see below) -- and that on a length scale of 10s of metres or larger water ice is present, which may increase the tensile strength by a factor 10, we argue that the desiccated (water-ice-free) dust should have a tensile strength between $\Sigma=0.3\ \mathrm{Pa}$ (water ice dominates the tensile strength in the measurements) and  $\Sigma=20\ \mathrm{Pa}$ (water ice is unimportant for the measured tensile strength). Thus, we get with Eq. \ref{Eq:tensilestrength} a range of pebble radii between $a_{\mathrm{min}}=0.0011\ \mathrm{cm}$ and $a_{\mathrm{max}}=0.59\ \mathrm{cm}$.

If the gas pressure at the bottom of the ice-free dust layer close to the surface of a comet nucleus exceeds the tensile strength, the whole layer is locally torn off and accelerated away from the comet. As long as the incoming solar heat flux is sufficiently large, the process of build-up and destruction of an ice-free pebble crust is repetitive and defines the dust activity of the comet. As the model cannot predict the lateral extent of the emitted dust layer, we can only speculate that the observation of free-flying dust with sizes of a few decimetres (see point 5 in this Section) is due to the ``folding'' of initially flat clusters of pebbles. The folding process is physically plausible, because the pattern of the gas flow around the cluster causes stresses within the cluster that can easily overcome the rolling resistance so that the cluster collapses to a spheroidal shape. Tensile-strength values of $\sim 1\ \mathrm{Pa}$ at the \textit{surface} of 67P are not in contradiction to measured \textit{bulk} values of 10-200 Pa \citep{hirabayashi2016}, because the increased lithostatic pressure towards the centre leads to an enduring increased tensile strength \citep{blum2014,blum2015,gundlach2016}. Given the size of the lobes of 67P and the increased stickiness of water ice \citep{gundlach2015} over the desiccated surface material, a range of 10-100 Pa is expected. On top of that, the gravitational strength towards the centre of 67P should be on the order of 100 Pa \citep{gundlach2016}.

Finally, it should be noted that the tensile-strength values measured on the surface of comet 67P (3-15 Pa; \citet{groussin2015}) are possibly an upper limit, because cliffs with lower tensile strengths would have collapsed at smaller sizes and to smaller fragments and, thus, may have escaped observations by OSIRIS.

\textit{7. Pebble-size constraint from direct observations by the CIVA camera on Philae.} Direct observation of resolved positive relief features with sub-cm spatial resolution was possible with the CIVA instrument on Philae. The 695 pebbles found on CIVA images \#3 and \#4 possess diameters between $3.7\ \mathrm{mm}$ and $16.25\ \mathrm{mm}$ \citep{poulet2016}. Their cumulative size distribution cannot be fitted with a single power law and is very steep at the large end and very flat at the small end. The size range within one standard deviation from the median of the 412 pebbles of the CIVA image \#4 is $6.0-10.6\ \mathrm{mm}$. Thus the pebble radii are between $a=3.0\ \mathrm{mm}$ and $a=5.3\ \mathrm{mm}$.

\subsection{Streaming instability criterion}

\textit{8. Pebble-size constraint from the streaming instability.} The streaming instability is capable of concentrating dust particles with Stokes numbers around $St \approx 0.1$ \citep{carrera2015,yang2016}. Larger particles possess larger stopping times so that the Stokes number is also a measure for the dust size. However, not all particle sizes are equally well concentrated. While concentration of pebbles with $St\approx 0.1$ is feasible for dust-to-gas ratios of $Z={\Sigma_{\mathrm{s}}}/{\Sigma_{\mathrm{g}}\approx 0.015}$ (where $\Sigma_{\mathrm{s}}$ and $\Sigma_{\mathrm{g}}$ are the surface densities of the solid and gaseous components of the protoplanetary disc), smaller and larger pebbles require higher metallicities, e.g., $Z\approx 0.02$ for $St \approx 0.006$ and $St \approx 0.6$ \citep{yang2016}. Assuming a MMSN model \citet{weidenschilling1977b}, the mid-plane gas density of the solar PPD at 30 au was $3\times {10}^{-13}\ \mathrm{g\ }{\mathrm{cm}}^{-3}$, the mid-plane temperature was 50 K, and the metallicity was $Z\approx 0.01$. Using published gas-drag equations \citep{weidenschilling1977a} and scaling the required higher dust-to-gas ratios of $Z=0.02$ by respective lower gas densities, caused, e.g., by partial dissipation of the solar nebula, we get pebble radii between $a_{\mathrm{min}}=3\times {10}^{-4}\ \mathrm{m}$ and $a_{\mathrm{max}}=0.03\ \mathrm{m}$ for which the streaming instability should work.

\subsection{How large are the pebbles forming comet 67P?}

It must be emphasized that different observations are sensitive to different pebble ``size'' definitions. While the temperature-sensitive devices (MUPUS-TM, MIRO) mostly react on the volume-to-surface ratio of the pebbles, i.e., linearly with pebble size, the tensile strength scales with pebble radius to the power of ${-2}/{3}$, the detection of dust by imaging is sensitive to the cross section of the particles, i.e. scales with radius squared, and the dust size distributions shown in Figure \ref{fig:09} use a variety of analyses and can mostly not distinguish between homogeneous dust aggregates and clusters thereof. We do not necessarily expect a straightforward agreement of all derived pebble radii if the true pebble sizes are non-monodisperse. No single observation should be over-emphasized (and over-interpreted) to constrain the formation process of comet 67P. However, a comparison between the different criteria (see Figure \ref{fig:02}) shows that a reasonable consensus can be reached among all measurements shown above for pebble radii in the range from $a\approx 3\ \mathrm{mm}$ to $a\approx 6\ \mathrm{mm}$, denoted by the hatched area in Figure \ref{fig:02}. It is interesting to note that this size range falls right into the minimum of the metallicity-Stokes number curve derived by \citet{yang2016} for the transition between non-SI and SI regimes of the PPD, assuming gas densities as in the minimum-mass solar nebula model. Thus, a metallicity of $Z \gtrsim 0.015$, only slightly raised above the solar value, is required.

\section{Discussion}

Using the above result that comet 67P likely formed by a gentle gravitational collapse of a bound clump of mm-sized dust pebbles, we will now make predictions about its properties and will show that these are in agreement with observations.

\subsection{\label{Sect:porosity}Porosity}

If we consider the structure of a planetesimal formed by the gentle gravitational collapse of a bound clump of pebbles, we can derive its volume filling factor \citep{skorov2012} and compare it to that of the nucleus of comet 67P. It must be noted that for a total mass of collapsing pebbles corresponding to a final planetesimal radius of 50 km or smaller, the mutual collisions of the pebble-sized dust aggregates are so gentle that they neither get further compacted nor fragmented \citep{wahlberg2017,wahlberg2014}. Thus, the volume filling factor of the pebble-sized dust aggregates is determined by the bouncing phase before transport to the outer solar nebula, which yields $\phi_{\mathrm{pebble}} \approx 0.4$ \citep{zsom2010,weidling2009}. In addition, the pebbles will be randomly packed throughout the volume of the planetesimal. There is insufficient gravitational stress to considerably deform or break individual aggregates inside the planetesimal (\citet{groussin2015}, but see also below). Hence, the volume filling factor of the pebbles has to fall between random loose packing (RLP) and random close packing (RCP) values \citep{onoda1990}. RLP is the loosest possible configuration under which low-cohesion spheres can pack and has a volume filling factor for low-gravity objects of $\phi_{\mathrm{RLP}}=0.56$ \citep{onoda1990}. As the gravitational stress inside comet 67P is on the order of $\sim 100$ Pa and, thus, higher than cohesive and shear strengths, the pebbles will always arrange in the densest possible configuration. Any external perturbations, such as vibrations or lithostatic stresses, will cause a further densification up to the RCP limit, which has been determined for equal-size spheres to be $\phi_{\mathrm{RCP}}=0.64$ \citep{onoda1990}. For polydisperse spheres, the RCP can reach somewhat higher limits, e.g. $\phi_{\mathrm{RCP}}=0.68$ for a log-normal size distribution with a standard deviation of 0.3 \citep{baranau2014}. Thus, we expect a total volume filling factor for the planetesimal between $\phi_{\mathrm{tot}}=\phi_{\mathrm{pebble\ }}\times \phi_{\mathrm{RLP}}=0.22$ and $\phi_{\mathrm{tot}}=\phi_{\mathrm{pebble\ }}\times \phi_{\mathrm{RCP}}=0.26\dots 0.27$ (or slightly higher for higher degrees of polydispersity among the pebbles). If the volume filling factor of the pebbles is $\phi_{\mathrm{pebble}}=0.46$, as recently inferred from Rosetta GIADA measurements \citep{fulle2016a}, we get $\phi_{\mathrm{tot}}=\phi_{\mathrm{pebble\ }}\times \phi_{\mathrm{RLP}}=0.26$ and $\phi_{\mathrm{tot}}=\phi_{\mathrm{pebble\ }}\times \phi_{\mathrm{RCP}}=0.29\dots 0.31$, respectively. Due to the high porosity contrast between pebbles and fractal aggregates, we here disregard that the void spaces among the pebbles are not empty but filled with fluffy dust clusters \citep{fulle2017b}.

The derivation of the total volume filling factor for comet 67P naturally depends on the assumed dust-to-ice ratio and material parameters (mass density, gas permittivity). In the literature, the inferred values are $\phi_{\mathrm{tot}}=0.25\dots 0.30$ (derived through gravity field measured by RSI; \citet{paetzold2016}), $\phi_{\mathrm{tot}}=0.21\dots 0.37$ (derived through direct density measurements of the pebbles by GIADA; \citet{fulle2016a}), and $\phi_{\mathrm{tot}}=0.15\dots 0.25$ (derived through the permittivity measured by CONSERT; \citet{kofman2015}), respectively. Taking into account the individual uncertainties and underlying assumptions, we can state that the most likely values for the volume filling factor are in the range $\phi_{\mathrm{tot}}=0.25\dots 0.31$. This is in excellent agreement with our prediction.

Thus, we can already conclude at this point that the bulk of the pebbles have never seen impacts with speeds greater than $\sim 10\ \mathrm{m}{\mathrm{\ s}}^{-1}$, because pebbles are being compacted to $\phi_{\mathrm{pebble}} \approx 0.55$ to $\phi_{\mathrm{pebble}} \approx 0.90$ (and structurally destroyed) for impact velocities of $50\ \mathrm{m}{\mathrm{\ s}}^{-1}$ and $1000\ \mathrm{m}{\mathrm{\ s}}^{-1}$, respectively \citep{beitz2016}. Even for the loosest possible pebble packing, RLP, this would result in $\phi_{\mathrm{tot}}=0.31\dots 0.50$, which exceeds the published values.

\subsection{\label{Sect:homogeneity}Homogeneity}

A gentle gravitational collapse of a bound clump of mm-sized pebbles leads to a homogeneous body on all size scales down to the pebble size. This is consistent with findings by CONSERT \citep{kofman2015} and RSI \citep{paetzold2016}, whose resolution limits are, however, much larger than the pebble size. The absence of scattering in the CONSERT data indicates the absence of large contrast in the dielectric constant within the investigated depth, suggesting that cavities or density enhancements larger or comparable to the wavelength (3.3 metres) do not exist within the volume examined. However, measurements by CONSERT and RSI cannot exclude voids in the $\sim 10-100$ m size range in the bulk of the comet nucleus, due to a lack of observational volume (CONSERT) and sensitivity (RSI). CONSERT data are compatible with a slow variation of the dielectric constant over the top $\sim$100 metres from the surface of 67P \citep{ciarletti2015}, which cannot be explained by our formation model and could well be an evolutionary process. At the current stage, the gravitational collapse of a  bound cloud of pebbles can also not account for the observed layering of the two lobes of comet 67P \citep{massironi2015}, which speaks for an individual formation (or evolution) of the two lobes.

Measurements based on observations by the OSIRIS instrument indicate a difference in mass density between the two lobes of 5-15\%, with the large lobe being denser \citep{jorda2016}. This could be the result of lithostatic compression inside the larger lobe of 67P. The lithostatic pressure inside a homogeneous object with radius $r$ and a mean mass density $\rho$ due to its own gravity at a depth $x$ is given by $p\left(x,r\right)=\frac{2}{3}\pi {\rho}^2 G \left[r^2-{\left(r-x\right)}^2\right]$. Here, $G$ is the gravitational constant. The volume-averaged pressure is $\overline{p}\left(r\right)=\frac{4}{15}\pi {\rho }^2Gr^2$, the peak pressure at the centre of the body is $p_{\mathrm{c}}(r)=\frac{2}{3} \pi {\rho }^2 G r^2$. For the radii of the two lobes of comet 67P, $r_1=1.43$ km and $r_2=0.89$ km, and densities ${\rho}_1 \approx 550\ \mathrm{kg\ }{\mathrm{m}}^{-3}$ and ${\rho }_2\approx 500\ \mathrm{kg\ }{\mathrm{m}}^{-3}$ \citep{jorda2016}, we get $\overline{p_1}=35$ Pa, $\overline{p_2}=11$ Pa, $p_{\mathrm{c,1}}=86\ \mathrm{Pa}$ and $p_{\mathrm{c,2}}=28\ \mathrm{Pa}$, respectively. Taking into account that the two lobes are in contact and that the centre of mass is inside the larger of the two lobes, the compressive stress inside the larger lobe is further increased. Its volume-averaged upper limit can be calculated by assuming that the whole body is spherical with a radius of $r=1.65$ km and a mass density of $\rho =530\ \mathrm{kg\ }{\mathrm{m}}^{-3}$ and yields $p\lesssim 100\ \mathrm{Pa}$.

Low-velocity impact experiments of glass spheres of 5 mm radius and 1.37 g mass (and, thus, a density of $\rho =2600\ \mathrm{kg\ }{\mathrm{m}}^{-3}$), dropped from $h=2$ mm above the flat surface, consisting of $\sim$20 layers of dust aggregates with $\sim$1 mm diameter, yielded intrusions of 1-2 mm \citep{isensee2016}. As the impact pressure $p_{\mathrm{imp}}=\rho g h \approx 25-50\ \mathrm{Pa}$ is constant on a scale length of the impactor radius and then decays inversely proportional to the distance squared \citep{beitz2016}, the resulting density increase was $\sim$10-30\%. As the lithostatic pressure contrast between the two lobes is on the same order as the impact pressure in the impact experiments, we expect a similar density contrast between the two lobes, which matches the findings by \citet{jorda2016}. Although impacts are dynamical processes, in this example the impact velocity of $0.14-0.20\ \mathrm{m\ }{\mathrm{s}}^{-1}$ is much smaller than the sound speed so that the impacts were quasi-static.

The compressive strength of overhangs on the 10 m scale on the surface of comet 67P were estimated to be 30-150 Pa \citep{groussin2015}. These strength values are reached within the large lobe, but not in the small lobe of comet 67P. Thus, we expect material failure only in the inside of the large lobe that leads to a densification, which is in qualitative agreement with the observation of the density contrast between the two lobes.

\subsection{Thermal inertia}

Our planetesimal-formation model can be readily used to derive the thermal inertia of the near-surface material of the nucleus of comet 67P. The thermal inertia is a measure of the ability of the material to resist changes in temperature due to energy fluxes, and is defined by $I=\sqrt{k \rho  c_{\mathrm{p}}}$, with $k$, $\rho$ and $c_{\mathrm{p}}$ being the heat conductivity, the mass density and the heat capacity, respectively. Using similar values as those in \citet{schloerb2015}, $\rho = 500 \mathrm{\ kg\ }{\mathrm{m}}^{\mathrm{-3}}$ and $c_{\mathrm{p}} = 500\ \mathrm{J\ }{\mathrm{kg}}^{-1}\ {\mathrm{K}}^{-1}$, and the model of the heat conductivity by \citet{gundlach2012} (see also Appendix A and Figure \ref{fig:07}) for $a = 5\mathrm{\ mm}$, i.e. $k = {10}^{-2}\ \mathrm{W\ }{\mathrm{m}}^{\mathrm{-1}}{\mathrm{K}}^{-1}$ for $T = 200$ K and $k = {10}^{-3}\ \mathrm{W\ }{\mathrm{m}}^{\mathrm{-1}}{\mathrm{K}}^{-1}$ for $T = 100$ K, we get thermal inertia between $I = 16\ \mathrm{J\ }{\mathrm{K}}^{-1}{\mathrm{m}}^{-2}{\mathrm{s}}^{-1/2}$ ($T=100$ K) and $I = 50\ \mathrm{J\ }{\mathrm{K}}^{-1}{\mathrm{m}}^{-2}{\mathrm{s}}^{-1/2}$ ($T=200$ K). This agrees very well with MIRO results of $I = 10-30\ \mathrm{J\ }{\mathrm{K}}^{-1}{\mathrm{m}}^{-2}{\mathrm{s}}^{-1/2}$ and $I = 10-50\ \mathrm{J\ }{\mathrm{K}}^{-1}{\mathrm{m}}^{-2}{\mathrm{s}}^{-1/2}$ \citep{schloerb2015,gulkis2015}.

\subsection{Outgassing of water vapour as a function of heliocentric distance and water-ice abundance on comet 67P}

Using the thermophysical model described in detail in Appendix A, we calculated the expected outgassing rates for water vapour as a function of heliocentric distance and compared it to values inferred from ROSINA measurements \citep{fougere2016}. The free model parameters are the pebble radius, the thickness of the desiccated dust layer and the sub-surface areal coverage by water ice. For not too small dust pebbles (see Figure \ref{fig:07}), however, the heat conductivity per pebble radius is constant for a given temperature so that the former two parameters degenerate to a single parameter, i.e. the depth of the desiccated dust layer in units of the pebble radius.

We determined vertical steady-state temperature profiles (as proxies for the diurnal and seasonal sub-surface profiles) for insolations ranging from the sub-solar point at perihelion to $10^{-4}$ that value (in 10000 steps) along the orbit of 67P and set the ice-dust boundary as a free parameter, ranging from the surface to 30 pebble radii deep. For each temperature profile, the temperature of the water ice at the ice-dust boundary and the corresponding sublimation rate were calculated (see Table 1 by \citet{gundlach2011b} for details). Due to the gas permeability of the covering dust-aggregate layers, only a fraction of the sublimating molecules can escape into space. The number of escaping molecules as a function of the thickness of the desiccated dust-aggregate layer is calculated using the ex
mental results obtained by \citet{gundlach2011b} (see Appendix A, Eqs. \ref{eq:s6}-\ref{eq:s5} and Table \ref{Table:01} for details) for a volume filling factor of the pebble packing of $\phi_{\mathrm{packing}}=0.59\approx \phi_{\mathrm{RCP}}$.

The outgassing rate of each surface segment is calculated by either choosing a thickness of two pebble radii, i.e. $x=2a$ in Eqs. \ref{eq:s6}-\ref{eq:s5}, (nominal model) or such a thickness $x$ that the pressure build-up at the ice-dust interface is maximal, i.e. $Z(T(x))$ in Eq. \ref{eq:s1} is maximal, which then leads to a corresponding outgassing rate (i.e., the amount of molecules able to penetrate through the covering material and able to escape into space). It turned out that both assumptions agree relatively well (the difference between the two methods in the outgassing rate per segment is less than 20\%, but the overall behaviour of the outgassing rate versus heliocentric distance follows the exact same trend) so that we here only follow the nominal model. To arrive at the total outgassing rate, the respective values of the individual surface segments are summed, taking the area of each surface segment and its time-dependent insolation into account. We assume pebbles with radii $a = 5\ \mathrm{mm}$ (see Figure \ref{fig:02}) and, hence, a thickness of the desiccated dust-pebble layer of $1\ \mathrm{cm}$.

\begin{figure*}
	\includegraphics[width=6.6cm, angle=90]{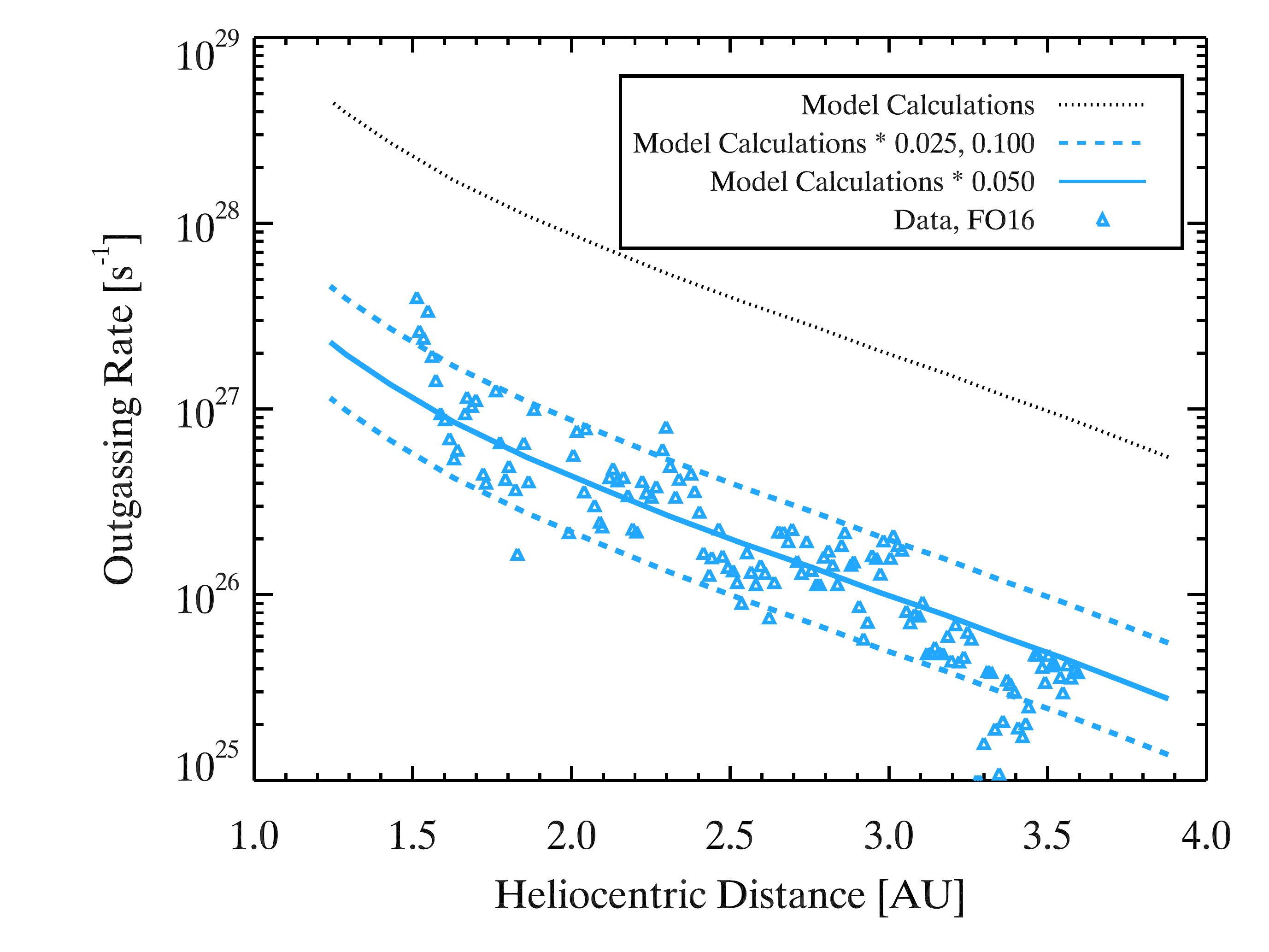}
    \includegraphics[width=6.6cm, angle=90]{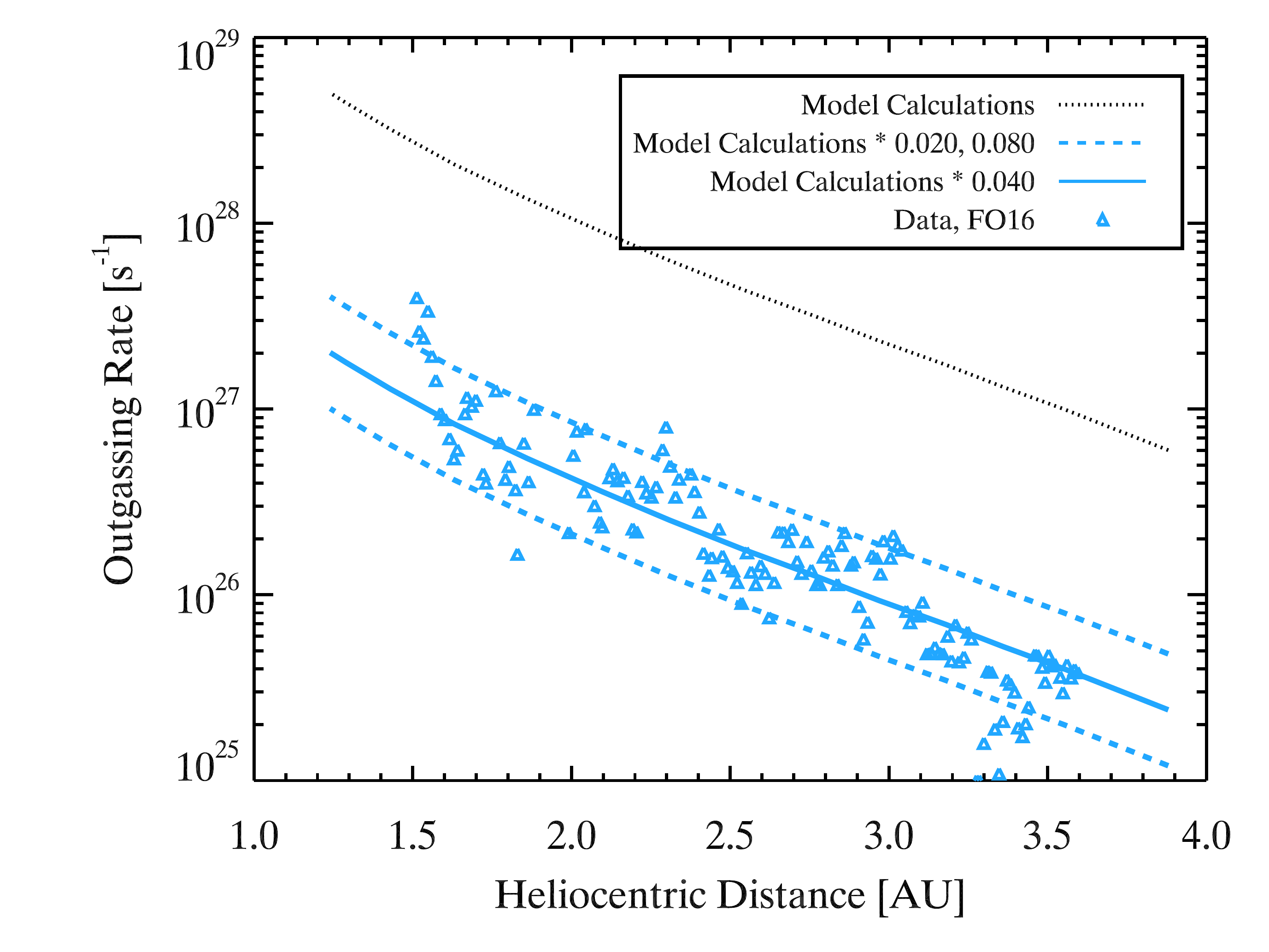}
    \caption{Water-vapour outgassing rates for comet 67P. Predictions of our thermophysical model (calculated for pebble radii of 5 mm) and an assumed dust-ice interface at 2 pebble radii (i.e. 1 cm) depth for the 67P outgassing rates of water ice (black dotted curves) and the measurements by Rosetta/ROSINA \citep{fougere2016} (triangles) are shown as a function of heliocentric distance. The solid and dashed blue curves denote models with different areal water-ice coverages at the dust-ice interface. Left: for a shape model of 67P consisting of 5000 facets and the obliquity of 67P and water-ice coverages of (from bottom to top)  2.5\%, 5\% and 10\%. Right: for a spherical comet consisting of 1000 facets with no obliquity and water-ice coverages of (from bottom to top) 2\%, 4\% and 8\%.}
    \label{fig:06}
\end{figure*}

We approximated the shape and orientation of 67P by a digital terrain model with 5000 facets \citep{jorda2016} and its real obliquity \citep{acton1996} as well as by a 1000-facet sphere with an effective radius of $1.65\ \mathrm{km}$ and a rotation axis perpendicular to the orbital plane. Figure \ref{fig:06} shows the resulting outgassing rates at different heliocentric distances derived for comet 67P. For comparison, the outgassing rates deduced from ROSINA data are shown by the blue triangles \citep{fougere2016}. The black curves present the results of the two models under the assumption that the entire surface is active and the ice-dust interface is completely covered in water ice.

A comparison with the measured outgassing rates provides the possibility to estimate the fraction of active surface area or the areal coverage by water ice. We find that both models (a comet with the shape and obliquity of 67P and a spherical comet, respectively) provide excellent fits to the measured data if we assume that $5\%$ of the dust-ice interface area is covered in water ice ($4\%$ for the spherical comet; see blue curves in Figure \ref{fig:06}). Thus, a homogeneous cometary sub-surface can explain the observed outgassing rate of comet 67P (see Figure \ref{fig:06}) and large areas of inactivity or hotspots of activity on the sunlit hemisphere are not required. This is in agreement with direct findings from Rosetta \citep{fulle2016b,kramer2016}. From Figure \ref{fig:06}, it can be assumed that an even better match to the ROSINA data, particularly at the highest and lowest heliocentric distances, can be achieved if the water-ice coverage at large heliocentric distances is slightly smaller than average, whereas for small heliocentric distances a somewhat larger areal water-ice coverage could be present. This dichotomy seems plausible, because regions active during the northern summer (at large heliocentric distances) are covered in fall-back material from the southern summer. The southern latitudes are active at small heliocentric distances and produce more material than they receive during the southern winter. As the fall-back material deposited in northern latitudes has been exposed to direct sunlight, we expect it to contain a somewhat smaller abundance of water ice (see \citet{fulle2017c}).

Some caution is required in interpreting the derived $5\%$ areal water-ice coverage or fraction of surface mass that is evaporating water ice. Assuming a thicker (thinner) desiccated dust cover leads to a diminished (increased) vertical heat transport that can formally be compensated by higher (lower) water-ice abundance. Thus, our model does not allow the prediction of the sub-surface water-ice coverage. However, if the water-ice abundance is known, we can use the model to estimate the thickness of the ice-free upper dust layer in units of the pebble radius.

With the assumption of an average dust-to-ice mass ratio of  $\delta =8.5$ \citep{fulle2016a}, we get a global water-ice mass fraction of $Y=\frac{1}{\delta +1}=0.105$ for the pristine comet nucleus. On the southern hemisphere, part of the water ice evaporates before the dust is ejected, leaving behind emitted dust with a water-ice mass fraction $Z$ and a mass fraction $X$ of the water vapour. Thus, we get $Y=X+Z$ for the southern hemisphere. For the northern hemisphere, we assume that 20\% of its surface is primitive \citep{keller2015}, with a water-ice content of $Y$, and 80\% of the surface is covered in deposit from the south, with a remaining water-ice abundance of $Z$. Therefore, we get for the northern hemisphere $0.2Y+0.8Z=X$, because the sublimating water content is about the same in the north and the south \citep{fulle2016b,fougere2016}. Solving the two equations with $Y=0.105$, we get $X=0.05$, in agreement with the needed mass surface fraction fitting ROSINA COPS data, and $Z=0.05$, in agreement with the ice mass content in dust derived by \citet{fulle2016b}. This consensus with the observations means that our assumption of a thickness of the desiccated dust layer of two pebble radii is consistent. These values are in accord with the ice abundances of other Kuiper-belt objects \citep{fulle2017a}.

It should also be mentioned that our outgassing model is not unique and that other models also predict a similar increase in water-production rate with decreasing heliocentric distance \citep{keller2015}. Thus, the evaporation rate of water ice is neither diagnostic for the pebble radii nor for the build-up of the desiccated dust crust. However, a correct prediction of the outgassing rate of water is a necessary condition for a comet-formation model.

\subsection{Size range for active comets}

Laboratory data indicate that the tensile strength of the interior of a body formed by pebble collapse increases towards the centre \citep{gundlach2016}. This is due to a memory effect of lithostatic pressure experienced at any time since formation. As the lithostatic pressure increases with increasing size of the body, large active comets should be scarce, because dust activity can only be present if the gas pressure exceeds the tensile strength. A comparison between (active) Jupiter-family comets (JFCs) and (inactive) asteroids in cometary orbits shows that the latter are on average larger than the former, although their perihelia distances are even smaller than those of the JFCs \citep{gundlach2016}. The size limit for activity was predicted to be between 2.7 and 4.5 km initial radius \citep{gundlach2016}. Both of the lobes of comet 67P are well below this limit so that they can sustain their activity until they vanish.

\section{Conclusion}

In conclusion, the analysis presented in this paper favours the formation of comet 67P through gravitational collapse of a bound clump of mm-sized dust pebbles and excludes any significant rearrangement (e.g. by catastrophic collisions) of the bulk of the pebbles since the gravitational accretion, which would have destroyed a significant fraction of the pristine fractals among the voids, rendering the observed rate of the GIADA showers of mm-sized fractal parent particles impossible \citep{fulle2017b}. However, even sub-catastrophic collisions as suggested by \citet{jutzi2017a} and \citet{jutzi2017b} cannot have acted on comet 67P. For an impact velocity of $v = 20 \mathrm{m \ s^{-1}}$ and near-equal-sized collision partners (as, e.g., in the case of the two lobes of comet 67P), we get a \emph{global} compression of $\sim \rho \ v^2 / 2 = 10^5$ Pa, with $\rho = 500 \ \mathrm{kg \ m^{-3}}$. This by far exceeds the tensile strength of the dust pebbles ($\sim 1-10$ kPa) as well as the compressive strength of the fractals, which get maximally compressed when colliding at $v = 0.7 \ \mathrm{m \ s^{-1}}$ \citep{blum2000b}. Thus, any collision above $\sim 1 ~ \mathrm{m \ s^{-1}}$ can be excluded for both lobes of comet 67P, which may have accreted from two cometesimals at collision speeds $< 1 \mathrm{m \ s^{-1}}$ \citep{jutzi2015}.

The fact that each of the two lobes of comet 67P is a pebble pile correctly predicts the comet's porosity, homogeneity and thermal inertia. With some additional assumptions of the water-ice abundance and the sub-surface depth of the dust-ice interface, the outgassing rate of 67P can also be reconstructed. On top of that, the dominance of the comet's total mass by dust pebbles predicts tensile strengths with which the dust activity can also be explained \citep{skorov2012,blum2014}. The presence of a distinct size scale on comet 67P, as shown in Figure \ref{fig:02}, together with the fragility of the cometary material, excludes a formation of 67P by mutual collisions among larger and larger building blocks in the solar nebula, as proposed by \citet{weidenschilling1997} and \citet{davidsson2016}.

\section*{Acknowledgements}

We thank Helena Schmidt for producing the RCP data for Figure \ref{fig:01}, Bj\"{o}rn Davidsson and Paul Weissman for critical reviews of the first draft of the manuscript, Sam Gulkis for providing original MIRO data, Mark Hofstadter and F. Peter Schloerb for discussion about the MIRO data, Harald Mutschke for data and discussions on dust opacities, Wlodek Kofman and the CONSERT team for insight into the CONSERT data, Ekkehard K\"{u}hrt for extensive discussions of the second version of the manuscript, J\"{o}rg Knollenberg for providing MUPUS-TM data, and Philipp Heinisch for providing the Philae housekeeping data.

J.B. acknowledges support by DFG project BL298/24-1 as part of the Research Unit FOR 2285. M.K. thanks the Deutsche Forschungsgemeinschaft (SFB963) for support. R.S. thanks the Austrian Research Promotion Agency (FFG) for financial support. T.M. and M.S.B. acknowledge funding by the Austrian Science Fund FWF P 28100-N36. T.M. acknowledges the Steierm\"{a}rkische Sparkasse and the Karl-Franzens Universit\"{a}t Graz for their financial support. S.F.G and C.S. acknowledge the financial support of UK STFC (grant ST/L000776/1; ERF). Rosetta is an ESA mission with contributions from its member states and NASA.




\bibliographystyle{mnras}
\bibliography{sample}




\appendix

\section{Thermophysical model of the cometary nucleus}

Owing to the low albedo of comet 67P/Churyumov-Gerasimenko, incoming solar irradiation is absorbed almost perfectly. Based upon OSIRIS images, a Bond albedo of only 1.2\% at $480\ \mathrm{nm}$ wavelength was derived (see Table 3 in \citet{fornasier2015}) so that 98.8\% of the energy delivered by the Sun is being converted into heat in the upper few micrometres of the surface. From here, heat is transported to the deeper layers by conduction through the network of dust particles and by (infrared) radiation from pebble to pebble \citep{gundlach2012}. Once the heat wave reaches the deeper ice-bearing layers, we assume that the heat flux is partially used to evaporate the ice. Details about the energy-transport mechanisms follow below. The emerging water molecules can either directly escape into the vacuum of space or hit one or more dust aggregates on their way out (see Figure \ref{fig:01}). Inter-molecular collisions are relatively unimportant and are here neglected, as is the heat transport by water vapour. Gas-dust collisions lead to an outward-directed momentum transfer from the gas molecules to the dust aggregates. We treat this exchange of momentum by calculating the steady-state gas pressure at the ice-dust interface. The gas permeability of the dust layer is a function of its thickness \citep{gundlach2012} and leads to the build-up of relatively higher pressures at the dust-ice interface for a higher number of dust-aggregate layers.

In order to derive the outgassing rate of comet 67P, we developed a one-dimensional thermophysical model, which solves the heat transfer equation at different positions below the surface of 67P (see Table \ref{Table:01} for an overview of the model parameters). For the surface material, we assume that the gravitational collapse of the bound pebble cloud has gently formed a surface layer composed of dust aggregates (see Figure \ref{fig:01}) composed of water ice and dust \citep{blum2014}. The water ice has retreated from the first surface layers due to the solar heat and is located below the ice-dust boundary (i.e., at these depths, the dust aggregates may also contain water ice, or aggregates composed of water-ice particles, are found in between the dust aggregates). Energy absorption occurs at the first pebble layer (surface layer) of the nucleus and is derived by the steady-state energy balance equation,
\begin{equation} \label{Eq:02}
I_{\odot }\ \left(\frac{r_H}{1\ \mathrm{au}}\right)^{-2}\ \left(1-A\right){\mathrm{cos} \vartheta \ }=\varepsilon \ \sigma \ T^4_S+\lambda \left(T\right){\left.\frac{\partial T}{\partial x}\right|}_{\mathrm{Surface}}.
\end{equation}
Here, $I_{\odot}$ is the solar constant, $r_H$ is the heliocentric distance, $A$ is the bond albedo, $\vartheta$ is the angle between the surface normal and the incident solar radiation, $\varepsilon$ is the emissivity, $\sigma$ is the Stefan-Boltzmann constant, $T_S$ is the surface temperature, $\lambda (T)$ is the temperature dependent heat conductivity of the porous surface material and $x$ is the depth beneath the surface \citep{gundlach2011b}.

The heat conductivity is given by \citep{gundlach2012}
\begin{equation} \label{Eq:03}
\lambda \left(T\right)=\frac{16}{3}\ \sigma \ T^3\ l\left(a\right)+H\left(a\right)\ {\lambda }_{\mathrm{pebble}}\left(T,a\right).
\end{equation}
The first term on the right hand side of Eq. \ref{Eq:03} describes the heat conductivity through radiation inside the pores of the granular material with $l(a)$ being the mean free path of the photons in the void space, which is a function of the pebble radius $a$ through \citep{skorov2011}
\begin{equation}
l\left(a\right)=e_{\mathrm{mfp}}\frac{1\ -\ \phi_{\mathrm{RCP}}\ }{\phi_{\mathrm{RCP}}}\ a, \label{eq:s2}
\end{equation}
with $\phi_{\mathrm{RCP}}$ and $e_{\mathrm{mfp}}$ being the volume filling factor of the pebble packing and a scaling factor, respectively. The second term on the right hand side of Eq. \ref{Eq:03} is the heat conductivity through the network of aggregates, with $H(a)$ being the Hertzian dilution factor for a granular material, which can be expressed by
\begin{equation}
H\left(a\right)=\ {\left[\frac{9\ \pi }{4}\frac{1-\mu^2_{\mathrm{pebble}}}{Y_{\mathrm{pebble}}}\ \frac{{\gamma }_{\mathrm{pebble}}}{a}\right]}^{1/3}\ f_1\ e^{f_2 \phi_{\mathrm{RCP}}\ },
\end{equation}
where $\mu_{\mathrm{pebble}}$ and $Y_{\mathrm{pebble}}$ are the Poisson ratio and Young's modulus of the pebbles, $f_1$ and $f_2$ are coefficients required to derive the influence of the packing structure on the heat conductivity and ${\gamma }_{\mathrm{pebble}}$ is the specific surface energy of the pebbles that can be formulated as
\begin{equation}
\gamma =\ \phi_{\mathrm{pebble}}\ {\gamma }^{5/3}_{\mathrm{grain}}{\left[\frac{9\ \pi \ (1-\mu^2_{\mathrm{pebble}})}{r_0\ Y_{\mathrm{grain}}\ }\right]}^{2/3}\ ,
\end{equation}
with ${\gamma }_{\mathrm{grain}}$, $r_0$ and $Y_{\mathrm{grain}}$ being the specific surface energy, the radius and Young's modulus of the monomer particles \citep{gundlach2012}.

Finally, ${\lambda }_{\mathrm{pebble}}\left(T,r\right)$ is the internal heat conductivity of the pebbles, given by
\begin{equation}
{\lambda }_{\mathrm{pebble}}\left(T,r\right)={\lambda }_{\mathrm{solid}}\left(T\right){\left[\frac{9\ \pi }{4}\frac{1-\mu^2_{\mathrm{grain}}}{Y_{\mathrm{grain}}}\ \frac{{\gamma }_{\mathrm{grain}}}{r_0}\right]}^{1/3}\ f_1\ e^{f_2 \phi_{\mathrm{pebble}}\ },\ \label{eq:s3}
\end{equation}
with $\mu_{\mathrm{grain}}$ and $\phi_{\mathrm{pebble}}$ being Poisson's ratio and the internal packing structure of the monomers inside the pebbles \citep{gundlach2012}. For the heat conductivity of the solid refractory grains, we used
\begin{equation}
{\lambda }_{\mathrm{solid}}=0.5\ \mathrm{W\ }{\mathrm{m}}^{-1}\ {\mathrm{K}}^{-1}, \label{eq:s4}
\end{equation}
to account for the high abundance of organic material.

\begin{table*}
\caption{\label{Table:01}Physical properties used in the thermophysical model.}
\begin{tabular}{|p{2.0in}|p{0.6in}|p{1.3in}|p{1.6in}|} \hline
{Properties} & {Symbol} & {Value} &  \\ \hline
\multicolumn{4}{|p{2in}|}{{Monomer dust grains}} \\ \hline
Radius & $r_0$ & $1.0\ \times \ {10}^{-6}\ \mathrm{m}$ & - \\ \hline
Young's modulus & $Y_{\mathrm{grain}}$ & $5.5\ \times \ {10}^{10}\mathrm{\ Pa}$ & \citet{chan1973}\\ \hline
Poisson ratio & $\mu_{\mathrm{grain}}$ & $0.17$ & \citet{chan1973} \\ \hline
Heat capacity & $C$ & $1.0\ \times \ {10}^3\ \mathrm{J\ }{\mathrm{kg}}^{\mathrm{-}\mathrm{1}}\mathrm{\ }{\mathrm{K}}^{\mathrm{-}\mathrm{1}}$ &  \\ \hline
Specific surface energy & ${\gamma }_{\mathrm{grain}}$ & $0.01\mathrm{\ J\ }{\mathrm{m}}^{\mathrm{-}\mathrm{2}}$ & \citet{heim1999} \\ \hline
\multicolumn{4}{|p{2in}|}{{Dust aggregates (``pebbles'')}} \\ \hline
Radius  & $a$ & $1.58\ \times \ {10}^{-4}\ \mathrm{m}$\newline $5.00\ \times \ {10}^{-4}\mathrm{\ m}$\newline $1.58\ \times \ {10}^{-3}\ \mathrm{m}$\newline $5.00\ \times \ {10}^{-3}\ \mathrm{m}$\newline $1.58\ \times \ {10}^{-2}\ \mathrm{m}$ & - \\ \hline
Volume filling factor & $\phi_{\mathrm{pebble}}$ & $0.4$ & \citet{weidling2009,guettler2010,zsom2010} \\ \hline
Young's modulus & $Y_{\mathrm{pebble}}$ & $8.1\ \times \ {10}^3\ \mathrm{Pa}$ & \citet{weidling2012} \\ \hline
Poisson ratio & $\mu_{\mathrm{pebble}}$ & $0.17$ & \citet{weidling2012} \\ \hline
\multicolumn{4}{|p{2in}|}{{Water-ice properties}} \\ \hline
Latent heat of sublimation & $\Lambda$ & $2.86\ \times \ {10}^6\ \mathrm{J\ }{\mathrm{kg}}^{\mathrm{-1}}$ & \citet{orosei1995}\\ \hline
Sublimation pressure (Eq. \ref{eq:s1}) & $a_1$\newline $a_2$ & $3.23\ \times \ {10}^{12}\ \mathrm{Pa}$\newline $6134.6\ \mathrm{K}$ & \citet{gundlach2011b} \\ \hline
\multicolumn{4}{|p{2in}|}{{Global properties}} \\ \hline
Thermal emissivity  & $\varepsilon $ & $1.0$ & (assumed) \\ \hline
Bond Albedo  & $A$ & $0.012$ & \citet{fornasier2015} \\ \hline
Volume filling factor of the dust-aggregate packing  & $\phi_{\mathrm{RCP}}$ & $0.6$ & -\\ \hline
Scaling parameter for mean free path (Eq. \ref{eq:s2}) & $e_{\mathrm{mfp}}$ & 1.34 & \citet{skorov2011}\\ \hline
Structure Parameter (Eq. \ref{eq:s3}) & $f_1$\newline $f_2$ & $5.18\ \times \ {10}^{-2}$\newline $5.26$ & \citet{gundlach2012} \\ \hline
Heat conductivity of refractory grains (Eq. \ref{eq:s4}) & ${\lambda }_{\mathrm{solid}}$ & $0.5\ \mathrm{W\ }{\mathrm{m}}^{\mathrm{-}\mathrm{1}}\mathrm{\ }{\mathrm{K}}^{\mathrm{-}\mathrm{1}}$ & (assumed) \\ \hline
Permeability parameter (Eq. \ref{eq:s5}) & $b$ & $13.85\ \times a$ & \citet{gundlach2011b} \\ \hline
Surface-to-volume ratio (Eq. \ref{eq:s6}) & $q$ & $1.2\ \times \ {10}^6\ {\mathrm{m}}^{\mathrm{-1}}$ & - \\ \hline
\end{tabular}
\end{table*}

\begin{figure}
	\includegraphics[width=6.4cm, angle=90]{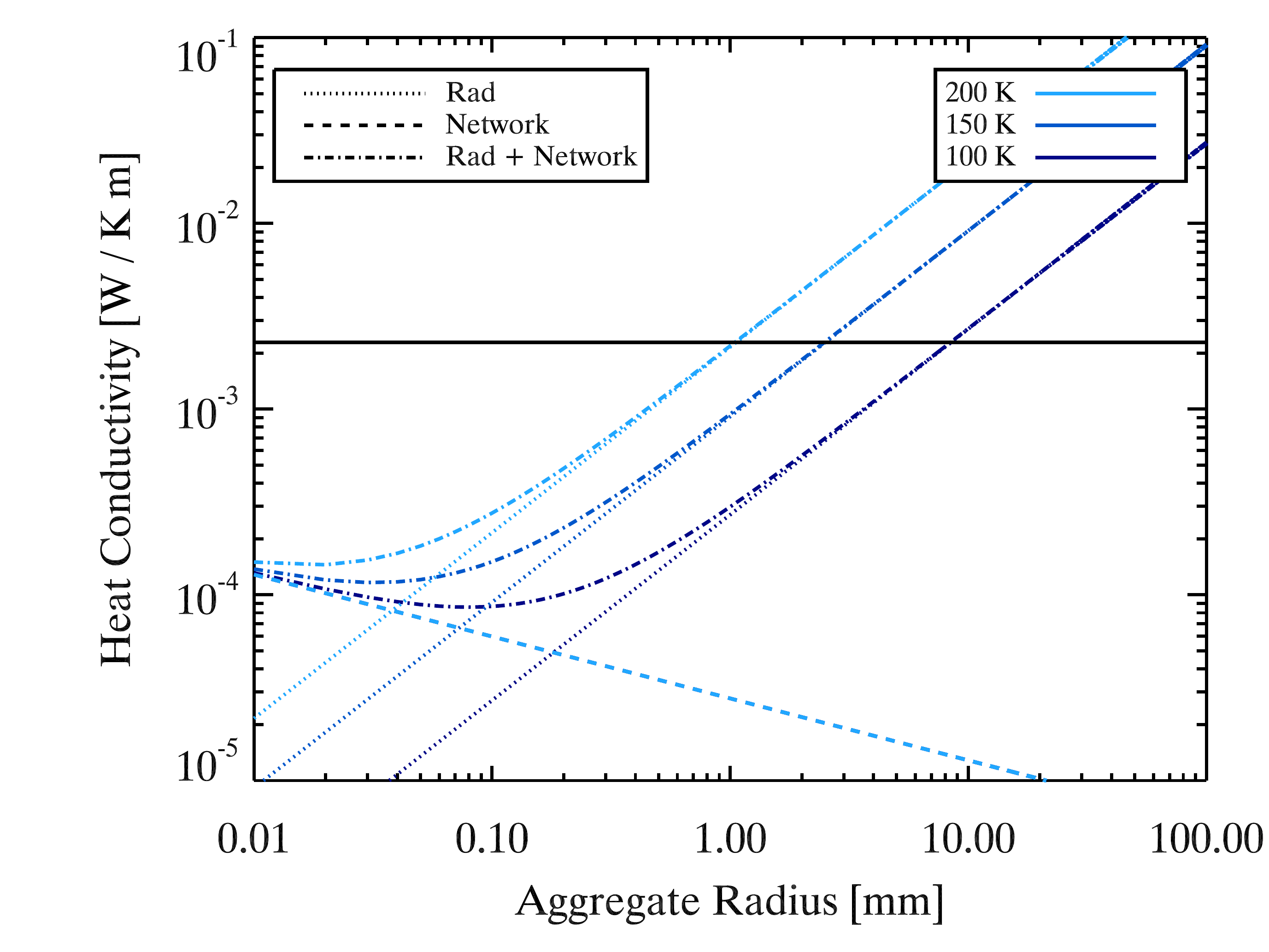}
    \caption{Heat conductivity as a function of dust-aggregate size. Shown is the heat conductivity of a network of spherical dust aggregates according to Eq. \ref{Eq:03} for temperatures of $T=100$ K,  $T=150$ K, and $T=200$ K, respectively. Due to the low solid-state heat conductivity through the pebble-pebble contacts (second term on the rhs of Eq. \ref{Eq:03}) of $10^{-5} - 10^{-4}$ $\mathrm{W}$ $\mathrm{m^{-1}}$ $\mathrm{K^{-1}}$, radiative heat transfer (first term on the rhs of Eq. \ref{Eq:03}) dominates for aggregate radii above $\sim$0.1 mm and all relevant temperatures. The horizontal black line at $2.2 \times 10^{-3}$ $\mathrm{W}$ $\mathrm{m^{-1}}$ $\mathrm{K^{-1}}$ denotes the heat conductivity through a homogeneous network of $\rm \mu$m-sized dust grains with individual heat conductivities of $\lambda_{\mathrm{solid}} = 0.5$ $\mathrm{W}$ $\mathrm{m^{-1}}$ $\mathrm{K^{-1}}$. }
    \label{fig:07}
\end{figure}

In Figure \ref{fig:07}, we compare the resulting heat conductivity according to Eq. \ref{Eq:03} for dust-aggregate radii between $a=0.01$ and $100\ \mathrm{mm}$. It turns out that radiative heat transport is the dominant effect for pebble radii above $\sim$0.1 mm. Thus, specific material properties, such as ${\lambda }_{\mathrm{solid}}$, are relatively unimportant for the heat transport through a network of pebbles and only the pebble radius determines the heat conductivity. It is interesting to note that the resulting heat conductivity for pebble radii of $\sim 1-10\ \mathrm{mm}$ and homogeneous dust layers consisting of $\rm \mu$m-sized monomer grains (horizontal black line in Figure \ref{fig:07}) is very similar. Thus, the heat conductivity (or thermal inertia) alone is not diagnostic for the absence or presence of pebbles.

At the start of each simulation, the initial temperature is set for all simulated depths $x$ to either $T=50\ \mathrm{K}$ or $T=133\ \mathrm{K}$, the latter being the orbital equilibrium temperature. The increase of the temperature below the surface due to the absorption of the solar radiation at the surface (Eq. \ref{Eq:02}) is derived by using the heat-transfer equation for porous materials,
\begin{equation} \label{Eq:09}
\rho c\frac{\partial T}{\partial t}=\ \frac{\partial }{\partial x}\ \left[\lambda \left(T\right)\frac{\partial T}{\partial x}\ \right]+S(T),
\end{equation}
where $\rho$ is the density of the porous material, $c$ is its heat capacity, $T$ is the temperature and $S(T)$ is an additional term, which takes the energy loss due to the sublimation process into account \citep{davidsson2002}, and is given by
\begin{equation}
S\left(T\right)=q\ Z\left(T\right)\ \Lambda\ \zeta \left(x\right). \label{eq:s6}
\end{equation}
Here, $q$ and $\Lambda$ are the surface-to-volume ratio and the latent heat of sublimation of water ice. $Z(T)$ is the sublimation rate, described by the Hertz-Knudsen equation
\begin{equation}
Z\left(T\right)=a_1\ e^{-{a_2}/{T}}\sqrt{\frac{m}{2\ \pi \ k_B\ T}} - Z_{0}, \label{eq:s1}
\end{equation}
where $a_1$ and $a_2$ are empirical constants describing the sublimation pressure of water ice, $m$ is the mass of a water molecule and $k_B$ is Boltzmann's constant. $Z_0$ describes the backflow of molecules onto the ice surface leading to recondensation. Normally, this term is known by $Z_{0} = p_{gas} \sqrt{\frac{m}{2\ \pi \ k_B\ T_{gas}}}$,
where $p_{gas}$ and $T_{gas}$ are the pressure and temperature of the gas phase above the ice surface. However, under cometary-like conditions, where molecules escape into space, $p_{gas}$ is practically zero so that this classical description of the backflow of molecules is not important for the energy budget of the system. However, the covering dust layer leads to backscattering of molecules onto the ice surface. The efficiency of this backscattering has been measured in laboratory experiments \citep{gundlach2011b}. Thus, we can formulate a new $Z_0$ describing this backscattering of water molecules by the covering dust layer onto the ice surface, which reads
\begin{equation}
Z_0 = Z(T) (1-\zeta \left(x\right)) .
\end{equation}
Here, $\zeta \left(x\right)$ is the fraction of escaping molecules as a function of the dust layer thickness $x$ and is given by
\begin{equation}
\zeta \left(x\right) \ = \ {(1+x/b)}^{-1},
\label{eq:s5}
\end{equation}
with $b$ being an empirical permeability parameter that depends on porosity (see Table \ref{Table:01}). With no dust cover ($x=0$), all molecules can freely escape into space and $\zeta \left(x=0\right) = 1$, i.e., $Z_0(x=0)=0$. For a pebble-layer with a thickness $x \gg b$, all molecules are effectively prevented from escaping from the comet nucleus and $Z_0(x \rightarrow \infty) \rightarrow Z(T)$.

The heat transport of the porous dust-aggregate layers strongly depends on the temperature of the material (see Figure \ref{fig:07}), because radiation inside the void space between the pebbles plays the dominant role in the energy transfer process \citep{gundlach2012} (see Eq. \ref{Eq:03} for details). We assume that the aggregates are primarily composed of dust \citep{fulle2016a,lorek2016}. This means that the dust determines the physical properties of the surface material (e.g., the network heat conductivity and the heat capacity). Water ice is incorporated into the model by allowing the material to sublimate at the ice-dust boundary (taken into account by the additional term $S(T)$ in the heat transfer Eq. \ref{Eq:09}). The position of the water-ice boundary beneath the surface is treated as a free parameter and is varied between $1$ and $30$ pebble radii to investigate how the positions of the ice-dust interface influence the resulting temperature profile.

Eq. \ref{Eq:09} is solved via the Crank-Nicolson method, where $S(T)$ on the right-hand side is treated as a superficial source term, if applicable \citep{hu2017}. The subsurface of the nucleus is discretized in depth into a number of numerical layers. The thickness of the layer, $\Delta x$, is always smaller than the pebble size in our simulation in order to resolve the fine variations of temperatures, especially near and at the surface. The upper boundary condition expresses the energy balance at the surface as given by Eq. \ref{Eq:02}. The isothermal condition is adopted for the lower boundary, such as

\begin{equation}
\left. \frac{\partial T}{\partial x} \right|_{x=\mathcal{X}} = 0\,,
\end{equation}
with $\mathcal{X}$ being the isothermal depth.

Because the energy input on the left-hand side of Eq. \ref{Eq:02} is periodic, it is mandatory to solve Eq. \ref{Eq:09} for the periodic variations of temperatures. At the starting epoch (which can be arbitrarily chosen for a certain heliocentric distance), say $t_0$, the solution is initialised, assuming that the nucleus subsurface is isothermal with a constant profile (either $T=50$ K or $T=133$ K throughout). The temperature profile is propagated in response to the varying energy input at each time step separated by $\Delta t$. To ensure numerical stability, the following criterion must be fulfilled

\begin{equation}
\Delta t \le \frac{c \rho}{\lambda}\frac{\Delta x^2}{2}\,.
\end{equation}

We perform the solution over precisely one comet rotation, for $t_0 \le t \le t_0+t_\mathrm{P}$, where $t_\mathrm{P}$ denotes the rotation period of 67P. The solution is iterated until the temperature profiles one rotation apart coincide, that is, $T(t_0) \approx T(t_0+t_\mathrm{P})$, which indicates convergence of the solution.

We note that the surface temperature is not provided directly by the Crank-Nicolson scheme. Instead, it can be solved from Eq. \ref{Eq:09} for the energy balance via the Newton-Raphson method at each time step.

\section{Modelling the MUPUS-TM data}

The above heat-conductivity model was additionally used to reconstruct the measured temperature data recorded by the thermal mapper MUPUS-TM on-board the Rosetta lander Philae during its stay on the landing site Abydos of comet 67P \citep{spohn2015}. From the housekeeping data of the solar cells on Philae, it can be seen that the solar cells received direct sunlight for about 40 minutes soon after the final touch-down. This duration is consistent with that of a temperature increase, recorded by MUPUS-TM, followed by a temperature decrease. We modelled the $\sim$28-minute temperature rise under the assumption that it was caused by insolation of the field of view of the thermal mapper. The modelling is based on the finite element method (FEM) algorithms served by the Partial Differential Equation Toolbox\textsuperscript{TM} of MATLAB (Release 2016b). For comparison of the measured temperature data with the results achieved by the numerical model, we used temperature values provided to us by the MUPUS team.

For the FEM modelling, the following simplifications were applied. The scenario to be modelled is a planar section of the comet surface heated from top by the Sun. As the pebble structure is assumed to be symmetric in the direction parallel to the comet surface, isothermal conditions hold and, thus, no heat exchange is present in this direction. The heat transfer by conduction from one pebble to another in the vertical direction is neglected in the model, because the contact regions are very small and few in comparison to the dominating amount of heat exchange by radiation between the pebble surfaces.

\begin{figure}
	\includegraphics[width=\columnwidth]{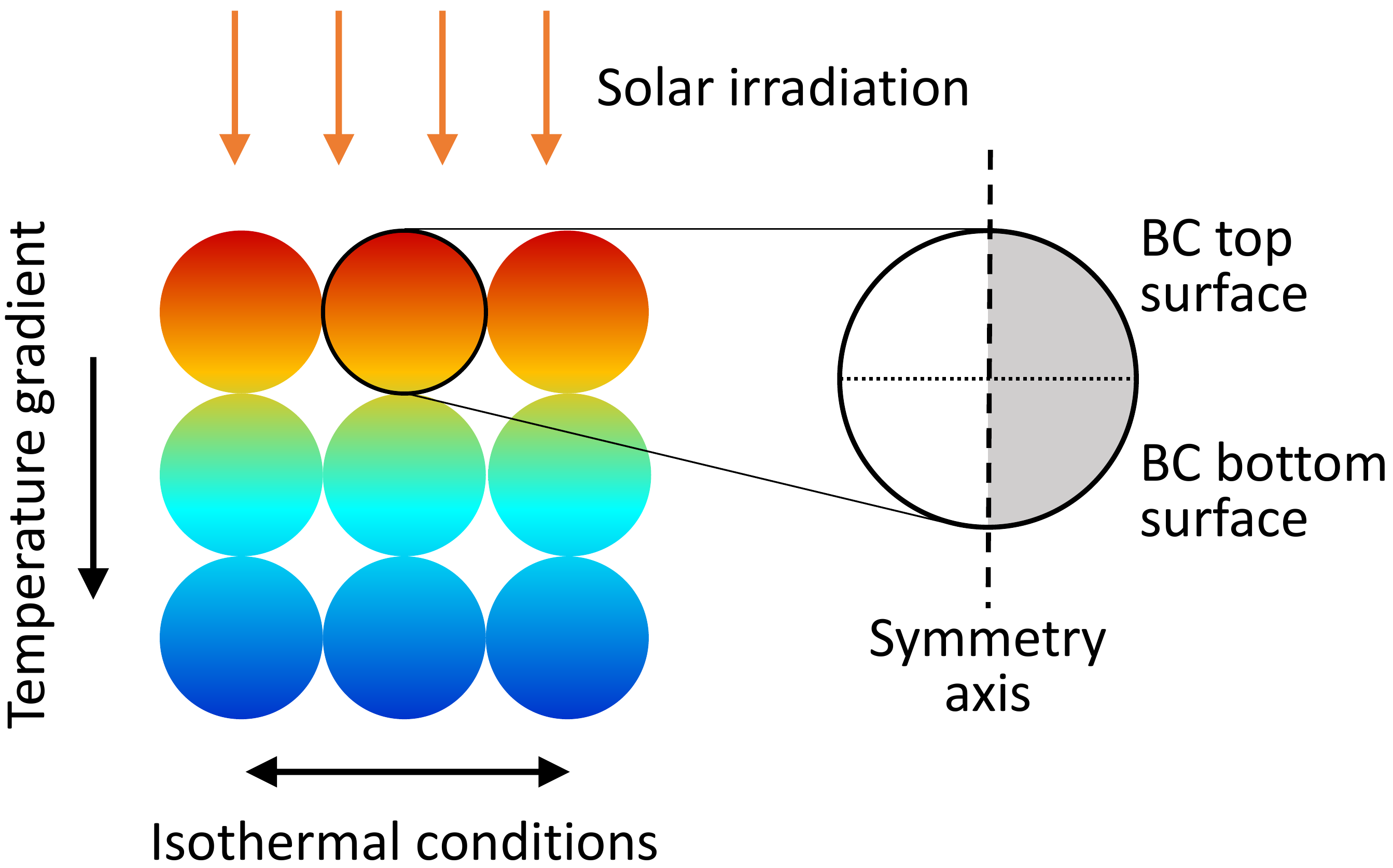}
	\caption{Sketch of the model setup for reproducing the MUPUS-TM temperature data. Left: section of the comet surface consisting of spherical pebbles irradiated from top by sunlight. Right: reduction of the heat transfer problem to a single surface pebble due to the constant temperature gradient in vertical direction and isothermal conditions in horizontal direction. The FEM model of the single pebble is calculated in cylindrical coordinates for a 2d semi-circle with the vertical centre line of the pebble as symmetry axis and appropriate boundary conditions (BC) for the top and bottom surface.}\label{fig:MUPUS-TM-FEM-model-sketch}
\end{figure}

Owing to these assumptions, the heat transport problem can be reduced to a single sphere representing a dust aggregate on the cometary surface, which is being irradiated at its top half by the Sun (see Figure \ref{fig:MUPUS-TM-FEM-model-sketch}).
Due to the rotational symmetry of the heat transfer inside the spherical pebble, the calculations could be reduced to a 2d model of a semi-circle with the symmetry axis at the centre line implemented in cylindrical coordinates. In the corresponding FEM model, following heat transport equation is solved
\begin{equation}\label{eq:heat-transport-cylcoord}
\rho \phi_{\mathrm{pebble}} c \frac{\partial T}{\partial t}  = \frac{1}{r} \frac{\partial}{\partial r} \left[ \lambda (T) \, r \, \frac{\partial T}{\partial r} \right] + \frac{\partial}{\partial z} \left[ \lambda (T) \, \frac{\partial T}{\partial z} \right],
\end{equation}
with the density $\rho \phi_{\mathrm{pebble}}$ of the porous pebble and its heat capacity $c$. The radial distance and height in the cylindrical coordinate system are given by $r$ and $z$, respectively. The effective heat conductivity is defined by
\begin{equation} \label{Eq:13}
\lambda \left(T\right)=\frac{16}{3}\sigma T^3l(r)+{\lambda }_{\mathrm{solid}}{\left[\frac{9\pi }{4}\frac{1-{\mu }^2_{\mathrm{grain}}}{Y_{\mathrm{grain}}}\frac{{\gamma }_{\mathrm{grain}}}{r}\right]}^{{1}/{3}}f_1e^{f_2{\phi }_{\mathrm{peb}\mathrm{ble}}},
\end{equation}
as the sum of heat radiation through the void spaces between the monomers (first summand) and heat conduction by the contact areas of the monomers (second summand). The radiative part of the effective thermal conductivity follows from Eq. \ref{Eq:03} adapted to a single pebble and the conductive part equals Eq. \ref{eq:s3}. For the chosen values in Eq. \ref{Eq:13}, refer to Table \ref{Table:01}. The initial condition $T_0$ for solving Eq. \ref{eq:heat-transport-cylcoord} was extracted from the temperature data provided by the MUPUS team.

The heat exchange of the pebble with the surroundings, i.e. the irradiation of sunlight, heat re-radiation and radiational heat exchange with neighbouring pebbles, is implemented as specific boundary conditions at the top and bottom surface of the model sphere. The heat exchange by radiation with underlying pebbles at the bottom half of the sphere is incorporated in the model as a radiation boundary with an ambient temperature representing the surrounding pebbles. To consider the local temperature evolution during the heating period at the bottom surface of the pebble, influenced by the temperature evolution of the interior of the comet, two limiting cases were calculated: (1) The ambient temperature at the bottom half of the sphere remains constant and equal to the initial temperature $T_{\mathrm{ambient}}=T_0$. (2) The ambient temperature at the bottom half of the sphere evolves like the temperature at the comet surface measured with MUPUS-TM. Both approaches correspond to the two extreme cases in which (1) no heat reaches the bottom of the pebble during the time of illumination, which is more likely for large pebbles, and (2) the heat is transferred from the top to the bottom of the pebble so efficiently that an isothermal condition is reached during the timescale of the insolation, which is the case for small pebbles.

At the centre axis of the model sphere, a zero Neumann boundary condition was applied. The boundary condition at the surface of the top hemisphere is defined according to Eq. \ref{Eq:02} by the balance of the absorption of energy from the Sun, the heat transport in the pebble interior and the re-radiation of heat from the surface. To account for a lateral distribution of pebbles with isothermal properties in the horizontal direction, the heat exchange by radiation with the environment (assumed at an ambient temperature of $T_{\mathrm{ambient}}$ = 0 K) was restricted to the vertical direction by the factor ${\cos \vartheta}$, i.e.
\begin{equation}
\lambda \left(T\right){\left.\mathrm{\nabla }T\right|}_{\mathrm{top\ surface}}=\left[\varepsilon \sigma \left(T^4_{\mathrm{ambient}}-T^4\right)+I_{\odot }\left(\frac{r_H}{1\ \mathrm{au}}\right)^{-2}\left(1-A\right)\right]{\cos \vartheta},
\end{equation}
with $\vartheta$ being the angle between the local normal to the surface and the vertical direction. The heliocentric distance $r_H$ was chosen to be 2.99 au, corresponding to the distance between the Sun and comet 67P during the landing of Philae.

\begin{figure}
	\centering	 \includegraphics[width=0.65\columnwidth]{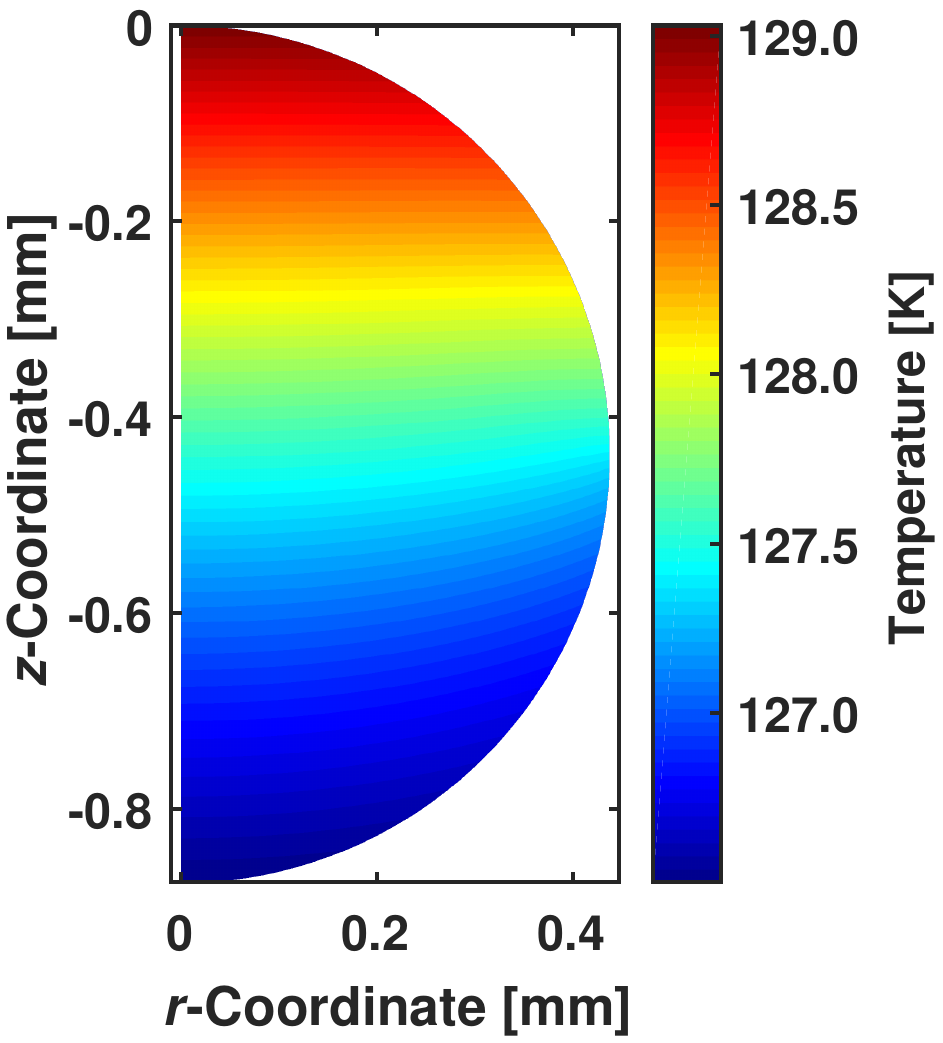}
	\caption{Temperature distribution inside the pebble sphere after $\sim$~28~min solar illumination as a result of the FEM model. The corresponding parameters for optimally fitting the measurements are  $a\ =\ 0.44\ \mathrm{mm}$ for the pebble radius and $f_{I_{\odot }}$= 0.16 for the illumination factor.}\label{fig:MUPUS-TM-FEM-model-result-temperature}
\end{figure}

For the bottom half-sphere, the boundary condition at the sphere surface is determined by the energy balance between the heat conduction of the pebble interior and the heat exchange by radiation at the surface with the ambient temperature representing the surrounding pebbles, i.e.
\begin{equation}
\lambda \left(T\right){\left.\mathrm{\nabla }T\right|}_{\mathrm{bottom\ surface}}=\varepsilon \sigma \left(T^4_{\mathrm{ambient}}-T^4\right).
\end{equation}
As the illumination angle is unknown, the solar constant $I_{\odot}$ is multiplied in the model by a factor $f_{I_{\odot}}$between 0 and 1, which is treated together with the radius of the pebble, $a$, as a free parameter. For comparison of the measured temperature at a given time of illumination with a single temperature value of the modelled sphere surface (see Figure \ref{fig:MUPUS-TM-FEM-model-result-temperature}), the (radiative) average temperature of the sphere surface was calculated by
\begin{equation}
T_{\mathrm{sphere\ surface}}={\left(\frac{\sum{d^2_iT^4_i}}{\sum{d^2_i}}\right)}^{{1}/{4}}.
\end{equation}
With this equation, the temperatures at the sphere surface at distances $d_i\ $from the centre axis of the 2d model are weighted by their squared distance to account for their corresponding surface ratio in 3d and are averaged by the fourth power to adopt the temperature measuring technique of the MUPUS-TM.

Finally, we also allow for partial direct illumination of the area observed by MUPUS-TM by applying the following mixing rule to the temperature:
\begin{equation} \label{Eq:17}
T_{\mathrm{model}}={\left({f_{\mathrm{s}}T}^4_{\mathrm{sphere\ surface}}+(1-f_{\mathrm{s}})T^4_0\right)}^{{1}/{4}}.
\end{equation}
The fraction of surface under direct illumination $f_{\mathrm{s}}$ was implemented in the model on the one hand as a constant free parameter and on the other hand as a time-dependent linear function $f_{\mathrm{s}}=f_{\mathrm{s,0}}+f_{\mathrm{s,1}}t$.

With the Fminsearch function in Matlab (Release 2016b), a function to find the minimum of a scalar function of several variables by the Nelder-Mead simplex search method, the optimal set of the pebble radius and the illumination factor can be found. The value to be minimised in this case is the mean squared difference of the MUPUS-TM temperature curve and the appropriate FEM model result
\begin{equation}
\left\langle {\Delta T}^2\right\rangle =\frac{\sum^N_{i=1}{{(T_{\mathrm{model},i}-T_{\mathrm{measurement},i})}^2}}{N}
\end{equation}
for the $N$ data points available in the heating curve from \citet{spohn2015}.

\section{Dust size distributions measured with Rosetta and from the ground}

An overview of the exponents $\beta$ of the size distributions (see Section \ref{Sect:RosettaData}) measured by various Rosetta instruments is given in Figure \ref{fig:09}a. COSIMA measured the size distribution from the material observed with the COSISCOPE camera on the collecting targets. The major unknown in this method is the degree of fragmentation the particles experience when hitting the target. The measurement near $\beta =-3$ is derived from pre-landing data and counts each particle individually \citep{hilchenbach2016}, while the remaining COSIMA results shown are based on measurements obtained up to April 2015 and take into account a medium degree of fragmentation \citep{merouane2016}. The COSIMA data shown here were averaged over several months. GIADA data are derived from combined measurements by the GDS+IS sensors during time spans of typically a few days. The value of $\beta =-2$ was measured at heliocentric distances beyond 2 AU, while $\beta =-3.7$ stems from perihelion \citep{rotundi2015,fulle2016c}. The OSIRIS coma data were derived from images of individual grains near the limb on four selected dates between August 2014 and August 2015 \citep{rotundi2015,fulle2016c}. OSIRIS surface data refer to boulder statistics in different terrains on the surface \citep{pajola2015}. The ROLIS data were obtained during Philae's approach to the Agilkia landing site and refer to ``rough'' (broken power-law) and ``smooth'' (single power-law) terrains, respectively \citep{mottola2015}. No indication that the size distribution measured in the centimetre to metre range continues down to millimetre and sub-millimetre sizes was found, as this would lead to a saturation of the surface with unresolved small particles, which is inconsistent with the observed granular texture. Therefore, the material must be depleted in small grains, consistent with a more shallow power-law at sub-centimetre-sizes. Ground-based measurements of the size distribution were obtained from numerical simulations of the morphology and brightness of dust tail and trail \citep{agarwal2010,fulle2010}.

It must be mentioned that the data presented in Figure \ref{fig:09}a does not distinguish between ``compact'' aggregates (which we interpret as pebbles or fragments/clusters thereof) and ``fluffy'' ones (which could be surviving solar-nebula aggregates, as suggested by \citet{mannel2016}, that formed contemporary to the formation of the planetesimals through gravitational collapse of a bound clump of pebbles; see \citet{fulle2017b}). Only GIADA is capable of deriving the mass density of the grains \citep{fulle2016c} by measuring cross section, momentum, and velocity of individual particles simultaneously. If we want to derive a joint size distribution function for all \textit{compact} aggregates, one has to be careful not to mix in data for the \textit{fluffy} particles. This is not a problem for the aggregates $\gtrsim 1\ \mathrm{cm}$ in size, because too large fluffy or fractal aggregates cannot survive between the (smaller) pebbles. However, those fluffy/fractal particles that fit into the void space between the $\sim 1$-cm-sized pebbles, must not be mixed with the pebbles themselves or fragments thereof. Thus, data from COSIMA has to be treated with caution, because they may contain contributions of both aggregate types.

The diagonal long dashed line in Figure \ref{fig:09}a is a fit to the data of the form $\beta =-0.48\ {\mathrm{log}}_{10}\ a-4.22$, with $a$ being the particle radius in metres, while the two other functions represent possible other, albeit extreme, linear approximations of the data to convey an impression of the uncertainty of the curves shown in Figure \ref{fig:09}b. Here, the normalized mass-frequency distributions per logarithmic size interval for the three linear approximations shown in Figure \ref{fig:09}a are shown. We can see that most of the mass is emitted in the form of decimetre particles. The strong decline in the mass-frequency distributions for sizes below $\sim$cm (or $\sim$mm for the curve labelled ``Extreme B'' in Figure \ref{fig:09}b) may indicate that this is the size of the primary building blocks (the ``pebbles'') of the comet nucleus. We interpret dust particles smaller than $\sim 1$ mm as pebble fragments due to the ejection process, and larger dust ``boulders'' ($\gg 1$ cm) as clusters of pebble-sized aggregates.


\bsp	
\label{lastpage}
\end{document}